\documentclass[10.5pt,letterpaper]{article}

\usepackage{natbib} 
\usepackage{fullpage} 
\usepackage{float} 
\usepackage{epsfig} 
\usepackage{amsmath} 
\usepackage{amsthm} 
\usepackage{amssymb} 
\usepackage{amstext} 
\usepackage{xspace} 
\usepackage{latexsym} 
\usepackage{verbatim} 
\usepackage{multirow} 
\usepackage{ifthen} 
\usepackage[hyphens]{url} 
\usepackage{cite}
\usepackage{aliascnt} 
\usepackage{enumerate} 
\usepackage[pdfpagelabels,pdfpagemode=None,breaklinks]{hyperref} 
\usepackage{graphicx}
\usepackage{subcaption}
\usepackage{authblk}
\usepackage{color,array}
\usepackage[width=.85\textwidth, font=footnotesize]{caption}
\usepackage{hyperref}
\hypersetup{
     colorlinks   = true,
     citecolor    = blue,
     breaklinks=true
}

 \theoremstyle{plain}

  \theoremstyle{remark}
  
  \theoremstyle{definition}
  
  \theoremstyle{plain}
  
\renewcommand{\cite}{\citep}

\newtheorem{theorem}{Theorem}[section]

\newtheorem{corollary}[theorem]{Corollary}
\newtheorem{remark}[theorem]{Remark}
\newtheorem{proposition}[theorem]{Proposition}
\newtheorem{example}[theorem]{Example}

\newtheorem{definition}[theorem]{Definition}
\newtheorem{assumption}[theorem]{Assumption}

\newtheorem{lemma}[theorem]{Lemma}

\newcommand{\ignore}[1]{}

\DeclareMathOperator*{\argmax}{argmax}

\newcommand*{\argmaxl}{\argmax\limits}

\makeatletter
\newcommand\footnoteref[1]{\protected@xdef\@thefnmark{\ref{#1}}\@footnotemark}
\makeatother

\newcommand*{\TitleFont}{%
      \usefont{\encodingdefault}{\rmdefault}{}{n}%
      \fontsize{16}{20}%
      \selectfont}
      
\newcommand{\Keywords}[1]{\par\noindent 
{\small{\\ \bf\em Keywords\/}: #1}}
     
\DeclareMathAlphabet{\mathcal}{OMS}{cmsy}{m}{n}

\begin{document}

\title{\TitleFont Fragility of the Commons under Prospect-Theoretic Risk Attitudes\thanks{Parts of this research were carried out when the authors were with the Department of Electrical and Computer Engineering, University of Waterloo, Canada. This research was funded by the Natural Sciences and Engineering Research Council (NSERC) of Canada under the Strategic Grants Program and by the Purdue Research Foundation (PRF). Preliminary versions of some of our results were presented at the \textit{Fifty-First Annual Allerton Conference on Communication, Control and Computing, 2013}~\cite{hota2013resource}.}}

\author[1] {Ashish R. Hota\thanks{E-mail: \texttt{ahota@purdue.edu}}} 
\author[2] {Siddharth Garg\thanks{E-mail: \texttt{sg175@nyu.edu}}} 
\author[1] {Shreyas Sundaram\thanks{E-mail: \texttt{sundara2@purdue.edu}}} 
\affil[1] {School of Electrical and Computer Engineering, Purdue University}
\affil[2] {Department of Electrical and Computer Engineering, New York University}
\renewcommand\Authands{ and }
\date{}
\maketitle

\begin{abstract}
We study a common-pool resource game where the resource experiences failure with a probability that grows with the aggregate investment in the resource. To capture decision making under such uncertainty, we model each player's risk preference according to the value function from prospect theory. We show the existence and uniqueness of a pure Nash equilibrium when the players have heterogeneous risk preferences and under certain assumptions on the rate of return and failure probability of the resource. Greater competition, vis-a-vis the number of players, increases the failure probability at the Nash equilibrium; we quantify this effect by obtaining bounds on the ratio of the failure probability at the Nash equilibrium to the failure probability under investment by a single user. We further show that heterogeneity in attitudes towards loss aversion leads to higher failure probability of the resource at the equilibrium.

\end{abstract}

\section{Introduction}
\label{section:introduction}
Common-pool resources (CPRs) are a broad class of shared resources characterized by two properties. First, they are non-excludable, meaning that it is (practically) infeasible to prevent any user from accessing them. Second, they are rivalrous or subtractable: higher use by one user leads to less availability for others. Selfish or myopic decision making by users competing for a CPR often results in suboptimal outcomes (including potential destruction of the resource) for the entire group; prominent examples include collapse of fish stocks due to overfishing~\cite{walker1992probabilistic} and global warming due to greenhouse gas emissions~\cite{ostrom1999revisiting}. In his seminal paper, Hardin popularized the phrase ``Tragedy of the Commons" to refer to such outcomes~\cite{hardin1968tragedy}. 

In this paper, we study a game-theoretic setting of a CPR where the resource experiences probabilistic failure due to overutilization. The possibility of resource failure leads to uncertainty in the outcomes for the players. In this context, the risk preferences of  the players can have a significant impact on their actions, and consequently on the utilization and fragility of the resource. Studies from behavioral economics show that individuals are typically neither risk neutral nor classical expected utility maximizers~\cite{von1947theory} when making decisions under uncertainty, and instead exhibit complex risk attitudes~\cite{machina1987choice}. One of the most widely accepted behavioral models of decision making under probabilistic uncertainty is ``Prospect Theory"~\cite{kahneman1979prospect,tversky1992advances}, where loss or gain of utility is measured with respect to a reference utility. Individuals exhibit risk seeking behavior under losses and risk averse behavior under gains, giving rise to an S-shaped utility function; this characteristic of the utility function is known as ``diminishing sensitivity." Furthermore, under prospect-theoretic preferences, the decrease in utility under loss of investment is typically greater than the increase in utility under a gain of the same magnitude; this behavior is popularly known as ``loss aversion." Thus, prospect-theoretic utility functions account for these systematic (and experimentally observed) deviations in human behavior from the predictions of the classical expected utility theory framework.

While prospect theory has been applied in diverse settings involving decision making under risk, including finance, insurance, industrial organization and betting markets (see~\cite{barberis2012thirty} for a review), theoretical analysis of prospect-theoretic preferences has been more recent and growing~\cite{butler2007prospect,baharad2008contest,leclerc2014prospect,easley2015behavioral}. As we discuss in Section \ref{ss:relatedwork}, certain prospect-theoretic characteristics such as framing effects, reference dependence and loss aversion have been observed, often in isolation, in past experimental studies related to CPR games. However, most of the (theoretical) investigations of CPR games have not considered prospect-theoretic risk attitudes, focusing instead on risk neutral and classical expected utility maximization behavior while modeling the risk preferences of human beings. Both of the above frameworks are typically more tractable to analyze compared to prospect theory. Understanding the effects of behavioral risk preferences in settings that model tragedy of the commons phenomena remains largely unexplored. Thus, given its strong behavioral foundations and experimental evidence from related settings, we model players' risk preferences according to the value function from prospect theory and study the effect of these risk preferences on the decisions made by players competing for a failure-prone CPR (where the uncertainty faced by each player arises from the potential failure of the shared resource). 

Our formulation builds upon well established game-theoretic models for CPR sharing~\cite{gardner1994rules,budescu1995common}. In the standard CPR game~\cite{walker1990rent}, players start with an initial endowment and choose their investments in two resources; one of the resources has a constant return on investment (safe resource), while the other is a CPR with a rate of return function (to be precisely defined in Section~\ref{section:problemformulation}) that is decreasing in the total investment in the resource.\footnote{~\citet{gardner1994rules} studied how individuals behave in controlled experiments for different variants of this standard CPR game, how their strategies compare to the outcome predicted by Nash equilibrium strategies, and what mechanisms lead to greater cooperation among players.}~\citet{walker1992probabilistic} studied probabilistic resource failure in a repeated CPR game, where the CPR could fail in any iteration with a probability that is a linearly increasing function of the aggregate investment by the players in that iteration. In their setting, players are assumed to be risk neutral and maximize the expected sum of utilities across a fixed and finite number of iterations. 

Failure of a shared resource has been explored in greater detail in a related game-theoretic setting referred to as the ``resource dilemma" problem~\cite{suleiman1988environmental,rapoport1992equilibrium,budescu1995common}. Here, the players participate in a single stage game where they choose their level of consumption from a resource of an unknown size (drawn from a prior probability distribution, typically taken to be the uniform distribution on a given interval). If the total consumption requested by the players is less than the size of the resource, the players receive their requested amount; otherwise, none of the players receive any benefit.~\citet{budescu1995common} model the risk preferences of players using the classical expected utility maximization framework, which captures risk averse (or risk seeking) behavior with a concave (respectively, convex) utility function.\footnote{The main focus of this line of work has been to study the effect of uncertainty (e.g.,~\cite{aflaki2013effect}), and game structure (such as the order of investments by the players~\cite{budescu1995positional}) on the equilibrium outcome.}

In the present work, we consider a single stage standard CPR game with resource failure. Players split their investments between a fragile CPR and a safe resource. If the CPR does not fail, each player receives a return that is proportional to her own investment and the rate of return of the CPR. If the CPR fails, the players receive no return from it. We model the failure probability of the CPR as an increasing and convex function of the total investment in the CPR. This includes failure probability functions that are linearly increasing within an interval, capturing the setting in resource dilemma problems \cite{budescu1995common} where the resource size is drawn from an interval uniformly at random. The convexity assumption is also motivated by the characteristics of many complex systems that undergo a sharp transition from one state to another with only a small change in the environment around a threshold, often referred to as a ``tipping point" of the system~\cite{gladwell2006tipping, lenton2008tipping,rockstrom2009safe,scheffer2009critical}. Convex failure probability functions are of interest since they can approximate a sharp transition of the resource from a safe state (one with a small failure probability) to a fragile state (marked by failure probability close to one) with a relatively small change in the underlying investment. 

Our analysis captures CPRs with both decreasing and increasing rate of return functions. Examples of resources with decreasing rate of return functions can be found in both natural and engineered systems such as fisheries, groundwater basins, forests~\cite{gardner1994rules} and communication/traffic networks~\cite{nisan2007algorithmic}. Such resources are said to exhibit negative externalities or ``congestion effects."\footnote{The CPR game formulation is related to a broad class of games known as ``congestion games"~\cite{rosenthal1973class}, and we discuss the relationship of that line of work to our setting in Subsection~\ref{ss:relatedwork}.} On the other hand, there are shared resources that benefit from more usage and/or a larger number of users due to positive externalities or ``network effects"~\cite{katz1994systems}. For example, increased usage of a peer to peer file sharing system, an operating system or an online gaming platform leads to better services and greater benefits to users~\cite{blumrosen2007welfare,johari2010congestible}. At the same time, users in online peer to peer file sharing systems often illegally distribute copyrighted material, and such websites (such as {\it Napster} for music and {\it Gigapedia} for ebooks), upon becoming large and well known, face crackdown by law enforcement agencies, leading to ``failure" of the resource~\cite{Wired2002,Verge2012}.\footnote{More generally, network effects can also lead to dominance of a single product over the entire market; this acts as an entry barrier for more efficient technologies and leaves users stuck with a potentially undesirable product due to high switching costs. The transition of the market towards a monopoly often exhibits ``tipping point" type behavior~\cite{farrell2007coordination}.} There is also evidence of increasing returns to scale in fisheries and other natural common pool resources~\cite{itam2014analysis,swallow1997eptd}. We study fragile resources with increasing rates of return as stylized models to capture settings similar to the above examples. We refer to our general game-theoretic formulation as a ``Fragile CPR game." 

As described above, we depart from the risk neutrality assumption of standard CPR games, and the classical expected utility maximization setting of resource dilemma games, and focus on understanding the impact of (potentially heterogeneous) prospect-theoretic risk preferences of the players on the utilization and fragility of the CPR.\footnote{Our formulation captures heterogeneous prospect-theoretic risk attitudes (with respect to both risk and loss aversion) across players. The degree of heterogeneity in loss aversion has been found to be significant in several experimental studies, e.g.,~\cite{gill2012structural} and~\cite{iturbe2011framing}.} In the context of the Fragile CPR game, we seek to answer the following questions:
\begin{enumerate}
\item Equilibrium characterization: does there exist a pure Nash equilibrium (PNE) in Fragile CPR games with prospect-theoretic players, and what are its characteristics? 
\item Effect of competition: how does the failure probability of the CPR at equilibrium under strategic investments by multiple self-interested players compare to that under optimal investment by a single user (i.e., when the resource is used as a ``private good"), and how is this affected by the prospect-theoretic risk preferences of the players and the resource characteristics? 
\item Impact of heterogeneity in risk preferences: which setting leads to larger failure probability, a society of users having homogeneous risk preferences or one with heterogeneous risk preferences?
\end{enumerate}

We give a brief overview of our main results, including answers to the above questions, below.

\subsection{Summary of results}

\subsubsection{Equilibrium characterization} 
We establish the existence and uniqueness of a PNE in Fragile CPR games when players have (potentially) heterogeneous prospect-theoretic risk attitudes, under certain general assumptions on the rate of return and the failure probability functions. In particular, our results hold for rate of return functions that are concave and monotonically decreasing or increasing, and failure probability functions that are increasing and convex in total investment. Under these assumptions, players' utility functions are not necessarily monotonic or concave. Our proof of existence of a PNE follows from a first-principles analysis of the best response correspondence. We further show that Fragile CPR games are instances of \textit{best-response potential games}~\cite{voorneveld2000best}, for which there are simple dynamics that allow all players to converge to the equilibrium. We refer to the failure probability of the CPR at the PNE of the game as the ``fragility" of the CPR.

\subsubsection{Effect of competition} 
To answer question (2) above, we first show that the total investment in the CPR at the PNE of a game with $n > 1$ homogeneous players is bounded within a multiplicative factor of the optimal investment by a single user, and that the bound increases as players become more risk averse in gains and risk seeking in losses as modeled by prospect theory. In order to compare the corresponding failure probabilities, we introduce the metric ``Fragility under Competition" (FuC), defined as the ratio of the fragility of the CPR with multiple users to the fragility with only a single user. 

The FuC increases as the failure probability function becomes increasingly more convex (captured by the family of functions $p(x)=x^\gamma$ with increasing values of $\gamma$, where $x$ is the total utilization level of the resource) and the transition from a safe state to a fragile state becomes sharper around a threshold ($x=1$). We identify different regimes under which FuC increases linearly versus exponentially in $\gamma$. One of the implications of our result is that for certain rate of return functions, the benefit (vis-a-vis reduction in fragility) of a central planner is greater when the planner accounts for prospect-theoretic risk preferences, as compared to the setting where players are assumed to be risk neutral. 

\subsubsection{Impact of heterogeneity in risk preferences}
Heterogeneity in prospect-theoretic preferences could be either in loss aversion or in the diminishing sensitivity of the utilities for large gains and losses. We show that all else being equal, heterogeneity in loss aversion leads to greater fragility of the resource at the PNE, compared to a game with homogeneous loss aversion indices (having the same value as the mean loss aversion index in the heterogeneous game). However, heterogeneity in the sensitivity parameter can result in higher or lower fragility, depending on the parameters of the game. 

\subsection{Related Work}
\label{ss:relatedwork}

\subsubsection{Prospect-theoretic preferences in experimental studies on CPR games}

Prospect-theoretic risk preferences such as framing effects, reference dependence and loss aversion have been observed in experimental studies on CPR games and related problems.

Cooperation levels among players in CPR games and public goods (PG) games with identical utilities have been observed to differ in experimental studies \cite{sell1997comparing,brewer1986choice,mccusker1995framing}, potentially due to framing effects and loss aversion. In the PG game, a player has to contribute from private wealth, which is perceived as a loss, while in the CPR game, a player utilizes a shared resource, which is perceived as a gain. Experiments in the above studies have shown that players cooperate more in the CPR game, consistent with players being loss averse. However, \cite{sell1997comparing} found that the difference in cooperation levels in both settings is negligible when communication among players is allowed. 

As discussed earlier, our CPR game formulation captures a class of resource dilemma problems considered in \cite{budescu1990resource,budescu1995common}. A closely related family of problems is threshold public goods games, where the public good is provided if the total contribution exceeds a (potentially uncertain) threshold. The effects of loss aversion, reference dependence and framing effects have been investigated in threshold PG games with deterministic threshold values in \cite{mccarter2010even} and \cite{iturbe2011framing}. In \cite{iturbe2011framing}, the authors estimate the loss aversion indices of the players, which were observed to lie in the interval $[0.88,1.7]$ with a mean $1.11$, exhibiting player-specific heterogeneity and a range of gain seeking and loss averse behaviors. The authors also studied the effect of probability weighting on users' strategies. 

In resource dilemma problems with uncertain resource size, the experimental literature is consistent with the observation that cooperation level decreases as uncertainty about resource size/threshold increases \cite{van2004we}. In particular, under greater probabilistic uncertainty about the threshold, individual resource utilization becomes more variable, likely due to individual differences in risk attitudes \cite{budescu1990resource}. In addition, increased uncertainty leads to overutilization of the resource in settings where the rate of return is constant and where the rate of return depends on the utilization levels of other users \cite{gustafsson1999overharvesting}. Our analysis on the effects of competition provides a theoretical characterization of the impact of behavioral risk preferences on resource utilization and fragility in resource dilemma problems with uncertain resource size. 

\subsubsection{Theoretical analysis of prospect-theoretic preferences in games}

Theoretical analysis of prospect theory and its attributes, such as reference dependence, loss aversion and probability weighting, to model decision making in multiplayer settings has been relatively recent. \citet{shalev2000loss} introduced ``loss aversion equilibrium" in strategic games where players' utilities are defined with respect to an endogenous reference point and players exhibit loss aversion. Building upon Shalev's work,~\citet{leclerc2014prospect} introduced several notions of equilibria in strategic games with finite action sets when players have prospect-theoretic risk preferences. His results show that equilibrium concepts defined under prospect-theoretic considerations often explain the observed behavior in experimental settings better than the risk neutral definition of PNE. This further motivates investigations of prospect-theoretic preferences in strategic multiplayer settings. 

Several recent papers have analytically examined the behavior of prospect-theoretic players in specific games with strategic uncertainty, outside of the setting considered in this paper. We mention some relevant ones that have a similar perspective as ours. The impacts of the probability weighting function and loss aversion have been examined in the contest theory literature in \cite{baharad2008contest} and~\cite{cornes2012loss}, respectively. Butler studies the impact of prospect theory in an ultimatum game~\cite{butler2007prospect}, while Eisenhuth~\cite{eisenhuth2012reference} studies revenue optimal mechanisms for selling a single item where players have endogenous reference points and are loss averse. \citet{easley2015behavioral} consider the problem of designing incentive mechanisms in online crowdsourcing environments where the users have prospect-theoretic risk attitudes. In our recent work \cite{hota2015interdependent}, we analyzed the impact of prospect-theoretic probability weighting in a class of interdependent security games. 

\subsubsection{Relationship with congestion games}

Our formulation builds upon the CPR game defined in~\cite{gardner1994rules}, which is an instance of an ``atomic splittable congestion game," a well studied model in the algorithmic game theory and networking communities~\cite{roughgarden2014local,cominetti2009impact,orda1993competitive}. While past literature has examined the existence~\cite{orda1993competitive}, uniqueness~\cite{bhaskar2009equilibria,richman2007topological} and inefficiency of equilibria~\cite{roughgarden2014local} for this class of games, there has been little work on resource failure and/or risk sensitive players in such settings. Only recently, Hoy considered risk averse players with concave utility functions and examined the existence of PNEs in atomic splittable congestion games with uncertainty in the cost functions~\cite{hoyconcavity}. On the other hand, congestion games with discrete strategy sets have been studied under resource failure \cite{penn2009congestion,penn2011congestion} and risk sensitive players~\cite{nikolova2013mean,piliouras2013risk,angelidakis2013stochastic}. However, these models have different mathematical structures than the atomic splittable congestion games with continuous strategy sets and prospect-theoretic players which are the focus of the present work.  

Resources with positive externalities have also been considered in the broad congestion game framework~\cite{blumrosen2007welfare}. In a related work~\cite{johari2010congestible}, the authors model systems that exhibit both network effects and congestion effects, where the former is dominant for low resource utilization, and the latter dominates as usage increases. Analogous to the above literature, we model resources with positive externalities via rate of return functions that are increasing in the total investments of the players.  

\section{Background and Problem Formulation}
\label{section:problemformulation}
In this section, we first give a brief overview of prospect theory and the standard CPR game~\cite{gardner1994rules}. We then formulate the Fragile CPR game and state the assumptions that we make about the parameters of the game. 

\subsection{Prospect theory}

Prospect theory was proposed by Kahneman and Tversky in their seminal paper~\cite{kahneman1979prospect} as a behavioral model of decision making under probabilistic outcomes. There are four main features of this model, (i) reference dependence: individuals evaluate uncertain outcomes with respect to a reference point, and exhibit different behavior for gains and losses; (ii) loss aversion: an individual experiences greater disutility in the event of a loss as compared to a gain of equal magnitude; (iii) diminishing sensitivity: the associated utility (value) function is concave for positive outcomes and convex for negative outcomes, indicating that individuals are \textit{risk averse} in gains and \textit{risk seeking} in losses; and (iv) probability weighting: individuals tend to overweight smaller probabilities and underweight larger probabilities. Individuals with prospect-theoretic preferences evaluate their expected utilities by multiplying the value function with weighted probabilities of outcomes. We discuss widely used parametrizations of such utility and probability weighting functions due to~\cite{tversky1992advances}.\footnote{A more comprehensive discussion of the forms of utility and probability weighting functions can be found in~\cite{wakker2010prospect},~\cite{tversky1992advances} and \cite{gonzalez1999shape}.} 

The utility for an (uncertain) outcome $z \in \mathbb{R}$ is of the form
\begin{equation}
u(z) =
\begin{cases}
(z-z_0)^\alpha & \text{when }z \geq z_0 \\
-k(z_0-z)^\beta & \text{otherwise},
\end{cases}
\label{eq:prospectvalue}
\end{equation}
where $z_0$ is the reference point with respect to which losses and gains are defined, and $\alpha,\beta$ are real-valued constants lying in the interval $(0,1]$.\footnote{We have excluded the possibility of $\alpha$ and $\beta$ admitting the value $0$ to maintain continuity of the utility function in our analysis.} Smaller values of $\alpha$ and $\beta$ result in greater sensitivity towards gains and losses of small magnitude compared to those of large magnitude. The parameter $k$ is a weighting factor that captures the sensitivity of players towards losses as compared to gains. When $\alpha = \beta$, an individual with $k>1$ weights her losses more than gains, which is termed as ``loss averse" behavior. Conversely, when $0 \leq k \leq 1$, players weight gains more than losses, and this behavior is termed as ``gain seeking"~\cite{wakker2010prospect}. Thus, for fixed $\alpha$, a larger value of $k$ implies a greater degree of loss aversion. In this paper, we capture both types of risk attitudes and let $k \in [0,\infty)$. We refer to $k$ as the \textit{index of loss aversion}, in keeping with standard terminology~\cite{kobberling2005index}. In this paper, we consider $\alpha = \beta$, and refer to $\alpha$ as the \textit{sensitivity parameter}.\footnote{The authors in~\cite{tversky1992advances} estimate that $\alpha = \beta$ from experimental data. Recently,~\citet{booij2010parametric} review the results of several experimental studies that estimate the parameters of the prospect theory value and weighting functions. They find that with the exception of a few studies, $\alpha$ and $\beta$ are always quite similar.} Note that when $\alpha=1$ and $k=1$, players are simply risk neutral (expected value maximizers).

Kahneman and Tversky also observed that unlike the classical expected utility maximization framework, individuals weight the utility of an uncertain outcome using decision weights that are different from the actual probability of the outcome~\cite{kahneman1979prospect}. In particular, individuals overweighted low probabilities and underweighted high probabilities. In their follow up work~\cite{tversky1992advances}, the authors proposed to compute the decision weights by weighting { \it cumulative} probabilities; i.e., a probability $p$ of obtaining at least (respectively, at most) a certain amount greater (respectively, less) than the reference point was weighted as
\begin{equation}
\pi(p) = \frac{p^\gamma}{(p^\gamma+(1-p)^\gamma)^{1/\gamma}}, 
\label{eq:prospectweighting}
\end{equation}
where $\gamma \in (0,1)$. The weighting function $\pi$ is concave for low probabilities, and convex for high probabilities, giving it an inverse S-shape. A smaller value of $\gamma$ results in a sharper increase in $\pi(p)$ when $p$ is close to $0$, capturing the observed overweighting of low probabilities. The proposed model in~\cite{tversky1992advances} is accordingly known as cumulative prospect theory.   

To lay the groundwork for a study of prospect-theoretic players in the resource sharing setting while keeping the analysis tractable, we focus on the effect of the value function~\eqref{eq:prospectvalue} on players' strategies and leave an investigation of the effect of probability weighting functions for future work. Note that the value function captures the first three features of prospect theory discussed above, and has similar characteristics in prospect theory and cumulative prospect theory. We allow players to have heterogeneous risk preferences, captured by their individual $\alpha$ and $k$ values.

\subsection{Common-pool resource game}

Consider the standard CPR game introduced in~\cite{gardner1994rules}. Let $\mathcal{N}=\{1,2,\ldots,|\mathcal{N}|\}$ be a finite set of players. Each player has an initial endowment $e$ which she must split between a CPR and a safe resource. We define the strategy of player $i$ as her investment in the CPR, denoted by $x_i$. The remainder of her wealth $e-x_i$ is invested in the safe resource. Thus, the set of feasible strategies for player $i$ is the closed interval $S_i = [0,e]$. Following standard notation, we denote $S=\prod_{i \in \mathcal{N}} S_i$ as the joint strategy space of all the players, and $S_{-i} = \prod_{j \in \mathcal{N}, j \neq i} S_j$ as the joint strategy space of all players except player $i$. 

The utility of a player is additive in the returns obtained from the safe resource and the CPR, with the former providing a constant return of $w$ per unit investment. The value generated by the CPR is defined in terms of a production function $F$ which depends on the total investment by all players in the CPR, denoted by $x_T = \sum_{i=1}^n x_i$. The utility a player receives from the CPR is proportional to her investment in the resource. Formally, let $\mathbf{x} \in S$ denote the vector of investments by the players in the CPR. The utility of player $i$ is a function $u_i:S_i \times S_{-i} \to \mathbb{R}$, given by
\begin{equation}
u_{i}(\mathbf{x}) =
\begin{cases}
we & \text{  if  } x_i = 0 \\
w(e-x_i) + \frac{x_i}{x_T} F(x_T) & \text{ otherwise. }
\end{cases}
\label{eq:CPRutilitydef}
\end{equation}
In~\cite{gardner1994rules}, the production function $F(x_T)$ is assumed to be concave, with $F(0) = 0, F'(0)>w$ and $F'(ne)<0$. Therefore, initial investment in the CPR is more attractive compared to the safe resource, while sufficiently high investment leads to a suboptimal outcome. Under these assumptions, the CPR game is an instance of a concave game~\cite{rosen1965existence}, which is known to possess a PNE. If the players are identical in their initial endowments, the game has a symmetric PNE. Furthermore, the total equilibrium investment in the CPR is higher than the level of investment that maximizes the social welfare, defined as the sum of individual utilities of the players. 

Note that when $F(x_T)$ is concave and $F(0) = 0$, the function $\frac{F(x_T)}{x_T}$ is non-increasing. We refer to this ratio as the \textit{rate of return function} of the CPR, defined as $r(x_T) \triangleq \frac{F(x_T)}{x_T}$. 

\subsection{Fragile CPR game}

A Fragile CPR game builds on the standard (single stage) CPR game by introducing the potential for resource failure. In particular, the CPR fails with a probability that is an increasing function of the total investment $x_T$. The failure probability function is denoted by $p(x_T)$. If the CPR fails, the players only receive the utility from their investments in the safe resource.

We model players with preferences towards uncertain outcomes governed by prospect-theoretic value functions defined in~\eqref{eq:prospectvalue}. We consider an index of loss aversion $k_i \in \mathbb{R}_{\geq0}$ and sensitivity parameter $\alpha_i \in (0,1]$ for each player $i$. We define the reference point $z_0$ of each player to be the value obtained when she invests her entire wealth in the safe resource, i.e., the reference utility is $we$. 

Let $x_i$ be the investment of player $i$ in the CPR as before. When $x_i = 0$, the wealth of player $i$ is exactly the reference utility, and therefore her reference-dependent utility is zero. Otherwise, her utility depends on the failure or survival of the CPR. If the CPR survives, her reference-dependent utility is proportional to $(x_i(r(x_T)-w))^{\alpha_i}$, which is obtained by subtracting the reference utility $we$ from~\eqref{eq:CPRutilitydef} and shaping the result according to~\eqref{eq:prospectvalue}. We normalize the rate of return function so that $w=1$ without loss of generality, and denote $\bar{r}_i(x_T) \triangleq (r(x_T)-1)^{\alpha_i}$ for notational convenience. When the CPR fails, the player obtains only the return of $w(e-x_i)$ from the safe resource. As this is less than the reference $we$, the reference-dependent utility for the player is given by~\eqref{eq:prospectvalue} to be $(-k_i x_i^{\alpha_i})$. 
 
We make some natural assumptions on the failure probability and rate of return functions for the subsequent analysis in the paper. We consider failure probability functions that are increasing in total investment with $p(\bar{x})=1$ for some $0 < \bar{x} < \infty$. In order to isolate the players' strategies from the impact of budget constraints, we assume that $\bar{x} < e$.\footnote{We leave the investigation of the effects of constrained endowments on overutilization and fragility of the CPR for future work. We discuss some of the consequences of removing this assumption in Section~\ref{section:conclusion}.} Without loss of generality, we consider $\bar{x}=1$. Any investment by a player with value greater than or equal to $1$ will result in certain failure of the CPR, which is undesirable for any player with a nonzero investment. Therefore, we restrict ourselves to the strategy set $S_i = [0,1]$ for each player (including $1$ to make $S_i$ a compact subset of $\mathbb{R}_{\geq 0}$). Thus we have $p:[0,|\mathcal{N}|] \to [0,1]$, with $p(x_T)=1, \forall x_T \geq 1$. 

As previously discussed, we will consider both the case where the rate of return is decreasing in total investment (exhibiting negative externalities or congestion effects as in~\cite{gardner1994rules}), and the case where it is increasing (exhibiting positive externalities or network effects as in~\cite{blumrosen2007welfare}), separately. Furthermore, we assume $r(x_T) > 1$ or equivalently $\bar{r}_i(x_T)>0$ for $x_T \in [0,1]$. Thus, unless the CPR fails, it always gives a higher return than the safe resource, i.e., the return from the CPR is a gain compared to the reference point.\footnote{The analysis for general monotone $r$ can be carried out in a region $[a,b] \subseteq [0,1]$ such that $r(x_T) \geq 1$, $\forall x_T \in [a,b]$. When $r$ is decreasing, we have $a=0, b \leq 1$, and otherwise $a \geq 0, b=1$.}  

We formally state our assumptions below. 

\begin{assumption} \label{assumption:CDCI}
We consider Fragile CPR games with the following properties.
\begin{enumerate}
\item The failure probability $p(x_T)$ is convex, strictly increasing and twice continuously differentiable when $x_T \in [0,1)$, with $p(1)=1$.
\item The player-specific rate of return $\bar{r}_i(x_T)$ is monotonic, concave, twice continuously differentiable and positive when $x_T \in [0,1]$ for each player $i \in \mathcal{N}$.
\item Player $i$'s strategy set is $S_i = [0,1], \forall i \in \mathcal{N}$.
\end{enumerate}
\end{assumption}

\begin{remark}
Note that for any rate of return function $r(x_T)$ which is increasing (respectively decreasing), concave, twice continuously differentiable, and greater than $1$ for $x_T \in [0,1]$, then for any $\alpha_i \in (0,1]$, $\bar{r}_i(x_T) = (r(x_T)-1)^{\alpha_i}$ is also increasing (respectively decreasing), concave, twice continuously differentiable and positive, and thus satisfies condition 2 in Assumption~\ref{assumption:CDCI}. For example, affine decreasing rate of return functions (widely considered in previous work on CPR games such as~\cite{walker1992probabilistic,keser1999strategic,apesteguia2006does,hackett1994role}) and constant rate of return functions with $r(x_T) = b > 1$ satisfy the above requirements.
\label{remark:concavityofrbar}
\end{remark}

\ignore{\begin{remark}
While the assumptions of Ostrom et al. regarding the production function 
Several past work on CPR games have considered quadratic production functions $F(x_T)$
These papers look at linear rate of return functions which satisfy our Assumption 1. (look for r greater than w in each)~\cite{keser1999strategic,apesteguia2006does,hackett1994role}
\end{remark}}

\begin{remark}
In the rest of the paper, we refer to a function $g: \mathbb{R} \to \mathbb{R}$ as increasing (decreasing) if for $x_1,x_2 \in \mathbb{R}, x_1 < x_2 \implies g(x_1) \leq (\geq ) g(x_2)$, and strictly increasing (strictly decreasing) if the inequalities are strict.
\end{remark}

We now formally define the prospect-theoretic utility of a player in the Fragile CPR game. Each player's utility is a function of her own investment and the aggregate investment by all other players. Accordingly, we define $\bar{S}_{-i}$ as the space of total investment by the players other than player $i$, i.e., $\bar{S}_{-i} = \{x \in \mathbb{R}_{\geq 0} | x=\sum_{j \in \mathcal{N},j\neq i} x_j, x_j \in S_j\}$. For $y_i \in \bar{S}_{-i}$, the utility of player $i$ is given by
\begin{equation}
u_{i}(x_i,y_i) =
\begin{cases}
(x_i(r(x_i+y_i)-1))^{\alpha_i} & \text{ with probability }(1-p(x_i+y_i)) \\
-k_ix_i^{\alpha_i} & \text{ with probability }p(x_i+y_i).
\end{cases}
\label{eq:utilitydef}
\end{equation}
The players are expected utility maximizers with respect to the utility function given by~\eqref{eq:utilitydef}. Thus, player $i$ maximizes
\begin{align}
 \mathbb{E}(u_{i}) &= (1-p(x_T))(x_i(r(x_T)-1))^{\alpha_i}+p(x_T)(-k_ix_i^{\alpha_i}) \nonumber
 \\ & = x_i^{\alpha_i}[\bar{r}_i(x_T)(1-p(x_T))-k_ip(x_T)] \nonumber 
\\ & \triangleq x_i^{\alpha_i}f_i(x_T).
\label{eq:expectedutility}
\end{align} 
We will refer to $f_i(x_T)$ as the \textit{effective rate of return} of player $i$. We denote a Fragile CPR game as $\Gamma(\mathcal{N},\{S_i\}_{i \in \mathcal{N}},\{u_i\}_{i \in \mathcal{N}})$, where the risk parameters $k_i$ and $\alpha_i$ are user-specific and are contained within the utility functions in~\eqref{eq:utilitydef}.

\section{Pure Nash Equilibrium: Existence and Uniqueness}
\label{section:existenseanduniqueness}
We start our investigation of Fragile CPR games by showing the existence and uniqueness of a PNE for the general case where players are completely heterogeneous in their risk preferences (i.e., have player-specific $\alpha_i$ and $k_i$ values). 

Let $B_i: \bar{S}_{-i} \rightrightarrows S_i$ be the best response correspondence of player $i$, i.e., $B_i(y_i) = \argmax_{x \in S_i} \mathbb{E}(u_i(x_i,y_i)), y_i \in \bar{S}_{-i}$. We drop the subscript $i$ in the following analysis for notational convenience as the results hold for every player $i$. We show that $B(y)$ is single-valued and continuous for $y \in \bar{S}$ where $\bar{S}=[0,|\mathcal{N}|-1]$, and monotonic in the total investment. Our main result follows from there. The proofs of all results in this section are given in~\ref{appendix:existenceproof}. 

Depending on $y \in \bar{S}$, any best response $b(y) \in B(y)$ of a player can be either to invest zero in the CPR (i.e., invest entirely in the safe resource), or to have a strictly positive investment with $0 < b(y) <1$. Note that $b(y)$ can never be 1, since that will cause the resource to fail with certainty, thereby yielding a nonpositive utility.\footnote{We assume that a player prefers to invest entirely in the safe resource if the maximum utility from the CPR in equation~\eqref{eq:expectedutility} is zero.} When $b(y)>0$, the best response investment satisfies the first order condition of the player's utility in~\eqref{eq:expectedutility}, given by 
\begin{align}
\frac{\partial \mathbb{E}(u)}{\partial x} &=x^\alpha f'(x_T)+\alpha x^{\alpha-1}f(x_T) \nonumber
\\ &= x^{\alpha-1} [xf'(x_T)+\alpha f(x_T)]=0, 
\label{eq:pnexi}
\end{align}
where $x = b(y)$ and $x_T = b(y)+y$. When $b(y)>0$, we must have $f(x_T) > 0$, because otherwise the player will receive a nonpositive payoff. Thus, $f'(x_T)<0$ for every player when she is playing a nonzero best response. The next lemma shows that all nonzero best responses for any given player must be such that the resulting total investment lies in a single interval of the strategy space. This is an important property which will be useful in proving several results later in the paper.

\begin{lemma} For each player in a Fragile CPR game, the following are true.
\begin{enumerate}
\item There exists a player-specific $\bar{y} \in [0,1]$ such that 0 is a best response if and only if $y \geq \bar{y}$.
\item When $\bar{y}>0$, $f(\bar{y})=0$ and there exists a player-specific interval $\mathcal{I} \subset [0,\bar{y})$ such that if $y < \bar{y}$, then all best responses for the player are positive, and each best response $b(y) \in B(y)$ satisfies $b(y) + y \in \mathcal{I}$.
\item For $x_T \in \mathcal{I}$, we have $f(x_T)>0$ and $f'(x_T)<0$. 
\end{enumerate}  
\label{lemma:interval}
\end{lemma}

\begin{figure}[t]
    \centering
    \begin{subfigure}[t]{0.5\textwidth}
	\centerline{\includegraphics[width=6.7cm,height=4.5cm]{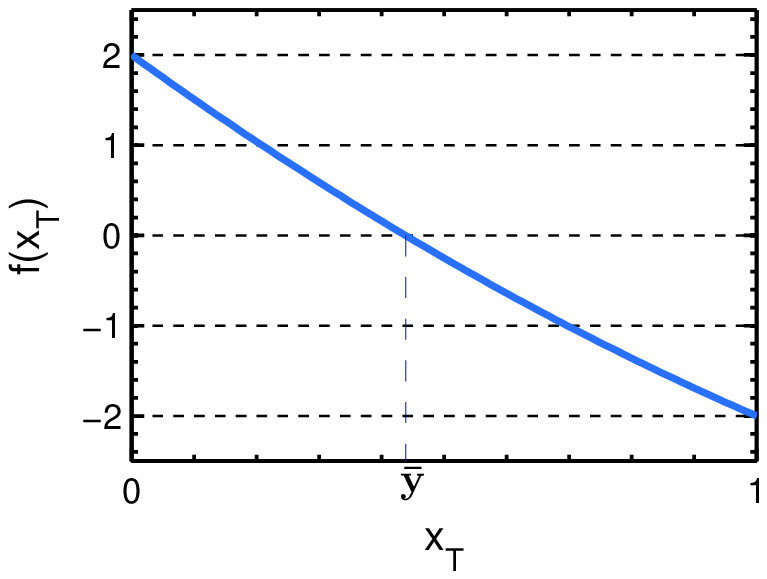}}
	\caption{Effective rate of return for decreasing $\bar{r}(x_T)$}
	\label{fig:lemmainterval1}
    \end{subfigure}%
    ~ 
    \begin{subfigure}[t]{0.5\textwidth}
	\centerline{\includegraphics[width=7cm,height=4.5cm]{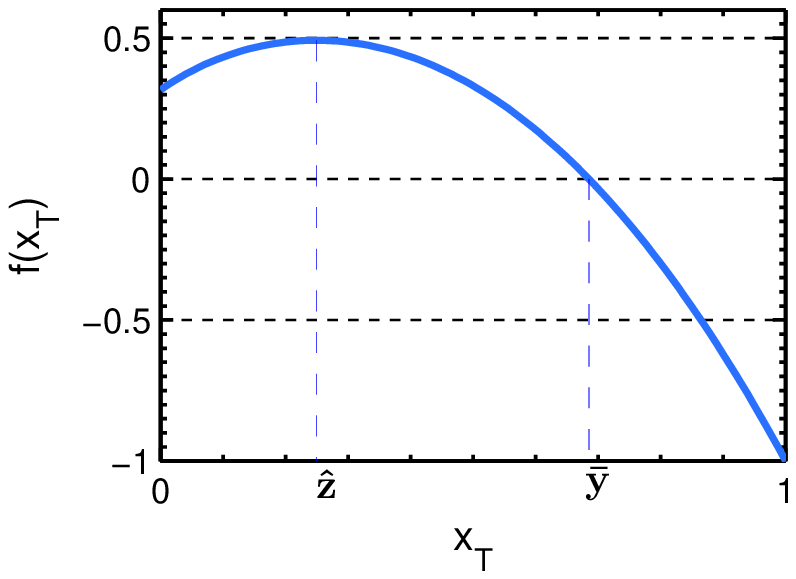}}
	\caption{Effective rate of return for increasing $\bar{r}(x_T)$}
	\label{fig:lemmainterval2}
    \end{subfigure}
    \caption{Illustration of the parameters $\bar{y}$ and $\mathcal{I}$ defined in Lemma~\ref{lemma:interval}: In Figure~\ref{fig:lemmainterval1}, we plot the effective rate of return $f(x_T)$ when $\bar{r}(x_T) = 2-x_T$, $p(x_T)=x_T$ and $k=2$. Note that $f(x_T)$ is decreasing in this case. We compute the quantity $\bar{y} = 0.4384$ which satisfies $f(\bar{y})=0$. The interval $\mathcal{I}$ is defined as $(0,\bar{y})$. Similarly in Figure~\ref{fig:lemmainterval2}, we plot $f(x_T)$ for $\bar{r}(x_T) = (x_T+0.1)^{0.5}$, $p(x_T)=x^2_T$ and $k=1$. In this case, we compute $\bar{y} = 0.6855$, which satisfies $f(\bar{y})=0$, and $f(x_T)<0$ for $x_T > 0.6855$. The required interval is $\mathcal{I} \triangleq (\hat{z},\bar{y})$, where $\hat{z} = 0.2493$ is such that $f'(\hat{z})=0$. Observe that $f(x_T)$ is concave in $x_T$, and for $x_T \in \mathcal{I}$, we have $f(x_T)>0$ and $f'(x_T)<0$.}
\end{figure}

We give the main idea behind the quantities $\bar{y}$ and $\mathcal{I}$ here, while relegating the formal analysis to~\ref{appendix:existenceproof}. Note that $\bar{y}$ and $\mathcal{I}$ are defined with respect to the (player-specific) effective rate of return function $f(x_T)$. When $\bar{r}(x_T)$ is decreasing, then $f(x_T)$ is also decreasing, and $f'(x_T) < 0$ for $x_T \in [0,1)$. If $f(0) \leq 0$, then any positive investment in the CPR can only do worse compared to zero investment. Therefore $\bar{y}=0$ in this case, and the interval $\mathcal{I}$ is undefined. Otherwise if $f(0) > 0$, there exists $\bar{y} \in (0,1]$ for which $f(\bar{y})=0$ and whenever $y < \bar{y}$, a player can find a nonzero best response investment in the CPR. Thus, the required interval is $\mathcal{I} = (0,\bar{y})$. We give a graphical illustration of $\bar{y}$ and $\mathcal{I}$ for this case in Figure~\ref{fig:lemmainterval1}. 

For increasing rate of return functions, $f(x_T)$ is not monotonic. However, we show that $f(x_T)$ is concave under Assumption~\ref{assumption:CDCI}, and there exists a unique $\bar{y}$ such that $f(\bar{y})=0$ and $f(x_T) < 0$ for $x_T > \bar{y}$. Furthermore, the set of total investments for which $f(x_T)>0$ and $f'(x_T)<0$ is an open interval contained in $[0,1]$, which we define as $\mathcal{I}$. We illustrate $\bar{y}$ and $\mathcal{I}$ with a numerical example in Figure~\ref{fig:lemmainterval2}. 

In the following two lemmas, we establish the uniqueness and continuity of the best response. 

\begin{lemma} The best response correspondence $B(y)$ is single-valued for $y \in \bar{S}$.
\label{lemma:CDuniquebr}
\end{lemma}

\begin{lemma} The best response correspondence $B(y)$ is continuous for $y \in \bar{S}$.
\label{lemma:CDcontinuity}
\end{lemma}

In the proof of Lemma~\ref{lemma:CDuniquebr}, we demonstrate the concavity of the utility function in a subset of the strategy set when $r(x_T)$ and $p(x_T)$ satisfy Assumption~\ref{assumption:CDCI}. We prove Lemma~\ref{lemma:CDcontinuity} using Berge's Maximum theorem (stated in~\ref{appendix:existenceproof}). The above two lemmas are sufficient to establish the existence of a PNE using Brouwer's fixed point theorem~\cite{ok2007real}. However, in order to prove uniqueness of the PNE, we first establish an important monotonicity property of the best response as follows. 

Notice from Lemma~\ref{lemma:interval} and Lemma~\ref{lemma:CDuniquebr} that there exists a unique nonzero best response to $y \in \bar{S}$ when $y < \bar{y}$. This best response $B(y)$ satisfies the first order condition in~\eqref{eq:pnexi}. Define 
\begin{equation}
g(x_T) \triangleq -\frac{\alpha f(x_T)}{f'(x_T)},
\label{eq:focbestresponse}
\end{equation}
which satisfies $g(B(y)+y)=B(y)$ when $B(y)>0$. In the next result, we observe that the optimal nonzero investment of a player, $g(x_T)$, is strictly decreasing in $x_T$. The proof follows by differentiating $g(x_T)$ and using the convexity and monotonicity properties of $r(x_T)$ and $p(x_T)$ in Assumption~\ref{assumption:CDCI}.

\begin{lemma}
The function $g(x_T)$ is a strictly decreasing function of $x_T$ when $x_T \in \mathcal{I}$, where $\mathcal{I}$ is the interval defined in Lemma~\ref{lemma:interval}. 
\label{lemma:garg}
\end{lemma}

The above lemmas lead to the following theorem.  In the proof of the theorem and the rest of the paper, we denote the unique Nash equilibrium strategy profile as $\mathbf{x}^* = \{x^*_i\}_{i \in \mathcal{N}}$ with total investment $x^*_T$. We define the ``support" of a PNE as the set of players with nonzero investment. Given the total investment $x^*_T$ at the PNE, the support is uniquely determined as 
\begin{equation}\label{eq:supportdef}
\mathcal{S} \triangleq \{i \in \mathcal{N} | x^*_T < \bar{y}_i\},
\end{equation}
where $\bar{y}_i$ is the player-specific threshold for investment identified in Lemma~\ref{lemma:interval}. 

\begin{theorem}
A Fragile CPR game admits a unique PNE. 
\label{theorem:pnecharacterization}
\end{theorem}

Note that since $p(x_T)=1$ for $x_T \geq 1$, we must have $x^*_T < 1$ at a PNE. Otherwise any player with a positive investment would prefer to invest in the safe resource and the strategy profile would not remain invariant under the best response mapping. 

\begin{remark} \textbf{Computing the best response:} Given the aggregate investment $y$ of all other players, computing the best response of a player involves finding the unique real root of a continuous and differentiable function as follows. To avoid trivial cases, we assume that the best response of the player is nonzero for some $y \in [0,1]$.  

From Lemma~\ref{lemma:CDuniquebr}, we know that the best response of a player is unique. This best response is zero if and only if $y \geq \bar{y}$ according to Lemma~\ref{lemma:interval}. If the best response is nonzero, it must satisfy the first order condition of optimality stated in~\eqref{eq:pnexi}. In other words, it must be a root of the function $l(x) \triangleq xf'(x+y)+\alpha f(x+y)$ in the interval $(0,1]$. 

In order to compute the best response, we first determine if it is zero, i.e., whether $y \geq \bar{y}$. For decreasing $r(x_T)$, this condition is satisfied if $f(y) \leq 0$ or equivalently $l(0) \leq 0$. On the other hand, for increasing $r(x_T)$, we have $y \geq \bar{y}$ if $f(y) \leq 0$ and $f'(y) < 0$. Therefore, for a given $y$, it is easy to check if $y \geq \bar{y}$ by evaluating the effective rate of return and its derivative at $y$, and conclude if the optimal investment in the CPR is zero.

If we find that $y<\bar{y}$, the best response is then the unique real root of $l(x)$ for $x \in (0,1]$, where $l(x)$ is a continuously differentiable function from Assumption~\ref{assumption:CDCI}. There are several numerical algorithms, such as the bisection method, that can compute the root of a continuous and differentiable function with any desired accuracy~\cite{fausett2003numerical}.
\end{remark}

\begin{remark} \textbf{Convergence to PNE:}
A direct consequence of our result in Lemma~\ref{lemma:garg} is that the best response of a player is decreasing in the total investment by all other players. As a result, Fragile CPR games belong to the class of games known as \textit{Weak Strategic Substitutes (WSTS) games with aggregation}~\cite{dubey2006strategic}. More generally, a strategic game is a WSTS game if there exists a continuous selection from the best response correspondence that is decreasing in the aggregate strategies of all other players. \citet{dubey2006strategic} define a continuous function of players' strategies and their best responses, and prove that in WSTS games, this function acts as a \textit{best-response potential function}~\cite{voorneveld2000best}. This has implications for the convergence of certain best-response dynamics to the PNE. We provide a brief discussion of this in~\ref{appendix:wsts}. 
\end{remark}

\section{Fragility Under Competition}
\label{section:fuc}
As discussed in the introduction, one of the goals of this study is to determine the impact of competition on the fragility of the CPR, i.e., how the failure probability of the resource shared among multiple players compares to the failure probability under private ownership. In order to quantify the increase in fragility of the resource utilized as a common property as opposed to a private property, we introduce the metric ``Fragility under Competition" (FuC). In order to isolate the effect of competition, vis-a-vis the number of players, on the fragility of the resource, we consider players with identical prospect-theoretic risk preferences (i.e., $k_i = k$ and $\alpha_i = \alpha$ for every player $i$). We refer to such players as homogeneous players. We obtain bounds on the FuC for the class of resources that satisfy Assumption~\ref{assumption:CDCI}. The effect of heterogeneous risk preferences on fragility is discussed in the next section. The proofs of all results in this section can be found in~\ref{appendix:fucproof}. 

We start with the formal definition of FuC. 

\begin{definition} \label{definition:FuC}
Consider a Fragile CPR game $\Gamma$ with $n \geq 2$ homogeneous players. Let the total investment at the PNE of $\Gamma$ be $x^*_{T}$, and the optimal investment by a single user (i.e., in the case where $n=1$) in the CPR  be $x_\mathtt{PVT}$. When $p(x_\mathtt{PVT})>0$, we define the \textit{fragility under competition (FuC)} of the game $\Gamma$ as
\begin{equation*}
\mathcal{F}(\Gamma) = \frac{p(x^*_{T})}{p(x_\mathtt{PVT})}.
\end{equation*}
On the other hand, when $p(x_\mathtt{PVT})=0$ and $p(x^*_{T}) > 0$, we define $\mathcal{F}(\Gamma) := \infty$. 
\end{definition}
Naturally, a game with a high FuC implies that the CPR under competition does relatively worse, vis-a-vis the fragility, with respect to the outcome when it is used as a private property. When $p(x_\mathtt{PVT})=0$, but $p(x^*_{T}) > 0$, the fragility is infinitely worse compared to when there is a single user. 

Without loss of generality, we will only consider games for which the effective rate of return for a player $i$ satisfies $f_i(x_T)>0$ for some $x_T \in [0,1)$, as this ensures that $x^*_{T}$ is nonzero irrespective of the number of players, and as a result, $p(x_\mathtt{PVT}) > 0$. Since players have identical risk preferences, we often omit the subscript $i$ in our analysis.

We begin by showing that fragility increases monotonically with the number of players, i.e., $\mathcal{F}(\Gamma) \geq 1$, and subsequently provide upper bounds on FuC in games with arbitrarily many players. 

\subsection{Monotonicity of investment and fragility with number of players}

Consider a Fragile CPR game with $n$ homogeneous players. In such a setting, all players participate in the PNE and have equal investment, which follows from the uniqueness of the PNE (Theorem~\ref{theorem:pnecharacterization}). We denote the total investment at a PNE with $n$ homogeneous players as $x^*_{Tn}$, with individual investment $\frac{x^*_{Tn}}{n}$. The investment of each player satisfies the first order condition of optimality (given by~\eqref{eq:pnexi}), stated as,
\begin{equation}
\frac{x^*_{Tn}}{n}f'(x^*_{Tn})+\alpha f(x^*_{Tn})=0,
\label{eq:homogeneous_foc_PNE}
\end{equation}
where $f$ is the effective rate of return for the common risk profile. The next result shows the behavior of $x^*_{Tn}$ as $n \to \infty$. The main argument of the proof applies our result on the monotonicity of the best response in the total investment stated in Lemma~\ref{lemma:garg}.

\begin{proposition}
Let $\{\Gamma_n\}_{n \geq 2}$ be a sequence of Fragile CPR games such that $\Gamma_n$ has $n$ homogeneous players with identical risk parameters $\alpha$ and $k$. Let the resource (i.e., $r(x_T)$ and $p(x_T)$) be fixed across all games in the sequence. Then the sequence of total investments at the PNEs, $\{x^*_{Tn}\}$, is strictly increasing and converges to $\bar{y}$, where $\bar{y}$ is as defined in Lemma~\ref{lemma:interval}. Consequently, the fragility of the CPR is strictly increasing in the number of players, and converges to $p(\bar{y})$.
\label{proposition:boundedpneXT}
\end{proposition}

Recall from Lemma~\ref{lemma:interval} that at $\bar{y}$, the effective rate of return satisfies $f(\bar{y}) = 0$. Therefore, unrestricted competition completely exhausts the utility derived from the shared resource, a characteristic often used to describe ``tragedy of the commons" phenomena. Proposition~\ref{proposition:boundedpneXT} is also useful in proving some of the subsequent results in this section.   

\subsection{Bounds on Fragility under Competition}

From Proposition~\ref{proposition:boundedpneXT}, we know that the total investment at the PNE for any game with the same resource and player risk characteristics as $\Gamma$ is upper bounded by $\bar{y}$, regardless of the number of players. Since $p(x_T)$ is increasing in $x_T$, we thus have $p(x^*_T) \leq p(\bar{y})$. Therefore, the FuC for any game is upper bounded as $\mathcal{F}(\Gamma) \leq \frac{p(\bar{y})}{p(x_\mathtt{PVT})}$; we will characterize this upper bound in the rest of the section. 

The main idea behind our proofs of upper bounds (for both decreasing and increasing rates of return) is the fact that for any given Fragile CPR game, there exists a perturbed game with affine $\bar{r}(x_T)$ such that the FuC of the perturbed game serves as an upper bound of the FuC of the original game. We formally state this in Lemma~\ref{lemma:gammahat} below. Our proof, presented in the appendix, exploits the concavity of the rate of return function. This approach is inspired by similar ideas used in~\cite{johari2004efficiency} to obtain bounds on the inefficiency of equilibria. 

\begin{lemma}
Given a Fragile CPR game $\Gamma$, there exists a perturbed game $\hat{\Gamma}$ with affine $\bar{r}(x_T)$ such that (i) the optimal investment by a single player remains identical for both $\Gamma$ and $\hat{\Gamma}$, and (ii) the quantity $\hat{\bar{y}}$ is an upper bound on $\bar{y}$, where $\hat{\bar{y}}$ and $\bar{y}$ are as defined in Lemma~\ref{lemma:interval} for $\hat{\Gamma}$ and $\Gamma$, respectively. 
\label{lemma:gammahat}
\end{lemma}

The consequence of this lemma is that the analysis can now be restricted to effective rate of return functions of the form $f(x_T) = (ax_T+b)(1-p(x_T))-kp(x_T)$, which are more tractable than the original form stated in equation~\eqref{eq:expectedutility}. Note that we have $a < 0$ (respectively, $a \geq 0$) when the rate of return is decreasing (respectively, increasing). 

We first obtain upper bounds on the ratio of total investment at the PNE with multiple competing users to the investment when the CPR is used as a private property. From the preceding discussion, it suffices to obtain a bound on the ratio of $\bar{y}$ and $x_{\mathtt{PVT}}$. We use the convexity of $p(x_T)$ to obtain these bounds. For both decreasing and increasing rates of return, the bounds depend only on the sensitivity parameter $\alpha$, and are independent of the loss aversion index and resource characteristics. The bounds on the ratio of investments lead to a bound on the ratio of failure probabilities (FuC) in terms of the failure probability function. We will focus on the behavior of the FuC bounds (and the actual FuC) in two distinct regimes of the failure probability functions: 
\begin{enumerate}
\item $p(x_T) = x_T$; this captures the setting considered in resource dilemma problems \cite{budescu1995common} where the resource fails to provide any return when the total demand for consumption exceeds a certain threshold, and the threshold is uniformly distributed over an interval (in this case $[0,1]$).
\item $p(x_T) = x_T^\gamma$; the CPR exhibits a sharp transition from a relatively safe state to a fragile state around the threshold $x_T = 1$ as $\gamma \to \infty$. This approximates a deterministic threshold in the limit.
\end{enumerate}

When $p(x_T)=x_T$, the bounds on the ratio of investments automatically act as bounds on FuC. When $p(x_T) = x_T^\gamma$, the behavior of FuC is qualitatively different depending on the maximizer of the function $x_T^\alpha \bar{r}(x_T)$ for $x_T \in [0,\infty)$. For strictly decreasing rate of return functions, the maximizer of $x_T^\alpha \bar{r}(x_T)$ is unique and is denoted as $x^*_r$. If $x^*_r < 1$, we obtain a lower bound on $\frac{p(\bar{y})}{p(x_\mathtt{PVT})}$ that grows exponentially in $\gamma$. On the other hand, if $x^*_r > 1$, we obtain an upper bound on FuC which grows linearly in $\gamma$. For increasing rate of return functions, the maximizer of $x_T^\alpha \bar{r}(x_T)$ is unbounded. Consequently our analysis yields an FuC upper bound which grows linearly in $\gamma$ in such cases.  

\subsubsection{Decreasing rate of return}

We start with an upper bound on the ratio of investments $\frac{\bar{y}}{x_\mathtt{PVT}}$. 

\begin{theorem}
Consider a Fragile CPR game $\Gamma$ with concave decreasing $\bar{r}(x_T)$,  convex strictly increasing $p(x_T)$, and $n \geq 2$ homogeneous players each having sensitivity parameter $\alpha$. Let $\bar{y}$ be the limit of the total PNE investments as $n \to \infty$, and $x_\mathtt{PVT}$ be the optimal investment by a single player in the CPR. Then we have
$$ \frac{\bar{y}}{x_\mathtt{PVT}} < \left(1+\frac{2}{\alpha}\right).$$ 
\label{theorem:FuCdecreasing}
\end{theorem}

As discussed above, the bounds arise from considering effective rate of return functions of the form $f(x_T) = (ax_T+b)(1-p(x_T))-kp(x_T)$ with $a<0$ and $b>0$. The proof relies critically on the monotonicity and convexity of the failure probability function. The above theorem shows that the total investment at the PNE under competition is always bounded within a constant multiple of the optimal investment under a single player. The bound is independent of the characteristics of the resource ($r$ and $p$), the number of players, and the index of loss aversion $k$, and only depends on the sensitivity parameter $\alpha$ for $\alpha \in (0,1]$. Note that when $\alpha = 1$ and $k=1$, the utility function is linear (leading to risk neutral behavior) and when $\alpha = 1$ and $k>1$, the utility function is piecewise linear and concave, thereby falling within the classical expected utility maximization setting. Thus as $\alpha$ decreases from $1$, players become more prospect-theoretic (increasingly risk seeking in losses and risk averse in gains) and this increases the bound on the ratio of investments. 

In the special case with $p(x_T) = x_T$ the above bound on the ratio of investments also serves as an upper bound on FuC. We now analyze the behavior of FuC for more general failure probability functions. As mentioned in the introduction, failure probability functions of the form $p(x_T) = x^\gamma_T$ (for large exponent $\gamma$) can approximate a sharp transition of the failure probability from a relatively safe state to a fragile state of the CPR, similar to the tipping point behavior observed in many complex systems. For this class of failure probabilities, the bound on investments in Theorem~\ref{theorem:FuCdecreasing} yields the following upper bound on FuC as a corollary. 

\begin{corollary}
Consider a Fragile CPR game $\Gamma$ with homogeneous players, concave decreasing $\bar{r}(x_T)$, and polynomial $p(x_T)$ with degree $\gamma$ and all coefficients nonnegative. Then we have $\mathcal{F}(\Gamma) \leq \left(1+\frac{2}{\alpha}\right)^\gamma$.
\label{corollary:FuCdecreasing}
\end{corollary}

Although the upper bound on FuC given in the result above grows exponentially with $\gamma$, the same is not always true of the actual FuC. In the following result, we show that the growth of FuC exhibits very different qualitative behavior depending on the maximizer of $x_T^\alpha \bar{r}(x_T)$ for $x_T \in [0,\infty)$. 

\begin{theorem}\label{theorem:fuclowerbound}
Consider a Fragile CPR game with concave strictly decreasing $\bar{r}(x_T)$ and homogeneous players each having sensitivity parameter $\alpha$ and loss aversion index $k > 0$. Let $p(x_T) = x^\gamma_T$. Let $x^*_r$ be the maximizer of $x_T^\alpha \bar{r}(x_T)$ for $x_T \in [0,\infty)$.
\begin{enumerate}
\item Suppose $x^*_r < 1$. Then, we have 
$$ \frac{\bar{r}(1)}{\bar{r}(1)+k} \left(\frac{1}{x^*_r}\right)^\gamma < \frac{p(\bar{y})}{p(x_\mathtt{PVT})}.$$ 
\item On the other hand, if $x^*_r > 1$, 
$$ \mathcal{F}(\Gamma) \leq \frac{p(\bar{y})}{p(x_\mathtt{PVT})} < \frac{\bar{r}(0)}{\bar{r}(0)+k} \left[1+\frac{\gamma(\bar{r}(0)+k)+k\alpha}{\bar{r}'(1)+\alpha \bar{r}(1)}\right].
$$
\end{enumerate}
In fact, the upper bound in the case where $x^*_r > 1$ holds more generally for any convex strictly increasing $p(x_T)$ with $\gamma$ replaced by the parameter $\zeta \triangleq \sup \left\{ \frac{x_Tp'(x_T)}{p(x_T)}, x_T \in (0,1) \right\}$ in the bound. 
\end{theorem}

Consider the case where $x^*_r<1$. Then the optimal investment by a single user ($x_\mathtt{PVT}$) is always less than $x^*_r$, since $x^*_r$ captures the optimal investment in the resource when $p(x_T)=0$. However, $p(\bar{y})$ remains bounded away from $0$ with a strict lower bound that is independent of $\gamma$. As a result, we obtain the above lower bound that is exponential in $\gamma$. Recall from Proposition~\ref{proposition:boundedpneXT} that the total investment at the PNE converges to $\bar{y}$ as the number of players $n \to \infty$. As a result, the above lower bound on $\frac{p(\bar{y})}{p(x_\mathtt{PVT})}$ also acts as a lower bound on FuC as the number of players $n\to\infty$.

On the other hand, if $x^*_r > 1$, then the optimal investment (and failure probability) for a single user approaches $1$ as $\gamma\to\infty$. As a result, the FuC grows only linearly in the exponent $\gamma$. The analysis relies on the fact that when $x^*_r > 1$, $x_T\bar{r}'(x_T)+\alpha r(x_T) > 0$ for $x_T \in [0,1]$. When $x^*_r = 1$, $\bar{r}'(1)+\alpha \bar{r}(1)=0$, and the provided upper bound is no longer finite.

\begin{remark}
Note that $\bar{y} \leq 1$ by definition. Therefore, $\frac{1}{p(x_\mathtt{PVT})}$ always gives us a trivial upper bound on FuC in terms of $x_\mathtt{PVT}$. However, this requires explicit computation of $x_\mathtt{PVT}$, while our bounds in Theorem~\ref{theorem:FuCdecreasing} and \ref{theorem:fuclowerbound} do not require computation of either $\bar{y}$ or $x_\mathtt{PVT}$. \end{remark}

In the following result, we obtain a necessary and sufficient condition for our bound in Theorem \ref{theorem:FuCdecreasing} to be tighter than the trivial bound $\frac{1}{p(x_\mathtt{PVT})}$ for $p(x_T)=x_T$.

\begin{proposition}\label{proposition:decreasing}
Consider a Fragile CPR game with concave decreasing $\bar{r}(x_T)$ and let $p(x_T) = x_T$. Then, $\left(1+\frac{2}{\alpha}\right)$ is a tighter upper bound on FuC compared to $\frac{1}{p(x_\mathtt{PVT})}$ if and only if 
$$ 2x_T\bar{r}'(x_T)+\alpha \bar{r}(x_T) \leq k\alpha(1+\alpha)$$
at $x_T = \frac{\alpha}{\alpha+2}$. In particular, if $\bar{r}(x_T) = (r(x_T)-1)^\alpha = ax_T+b$, then the above condition is equivalent to $\bar{r}(1) = a+b \leq k(1+\alpha)$. 
\end{proposition}

At $x_T = 1$, the resource experiences certain failure. For fixed $k$ and $\alpha$, and $\bar{r}(x_T) = ax_T+b$, as long as the rate of return function is sufficiently small at $x_T=1$, the bound obtained in Theorem~\ref{theorem:FuCdecreasing} is a stronger bound. If $\bar{r}(1)$ is large, i.e., the return from the CPR is large even when it has a very high failure probability, the optimal investment by a single user tends to be high, and as a result, $\frac{1}{p(x_\mathtt{PVT})}$ yields a stronger bound. 

While we do not have an analytical characterization of the strength of our bounds in Theorem \ref{theorem:fuclowerbound}, in the following numerical examples we show that $\frac{1}{p(x_\mathtt{PVT})}$ exhibits similar qualitative behavior as our analytical bounds in both regimes where $x^*_r < 1$ and $x^*_r > 1$. 

\begin{figure}[t!]
    \begin{subfigure}[t]{0.5\textwidth}
    \centering
        \includegraphics[width=7.8cm,height=5cm]{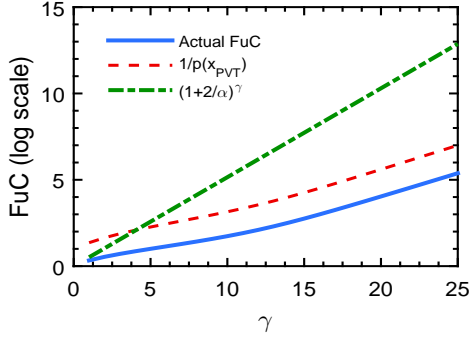}
        \caption{Exponential growth of FuC with degree $\gamma$ for \\ $p(x_T) = x_T^\gamma$ and $r(x_T) = 1.21-0.2x_T$}
        \label{fig:decreasing1}
    \end{subfigure}%
    ~ 
    \begin{subfigure}[t]{0.5\textwidth}
        \centering
        \includegraphics[width=7.8cm,height=5cm]{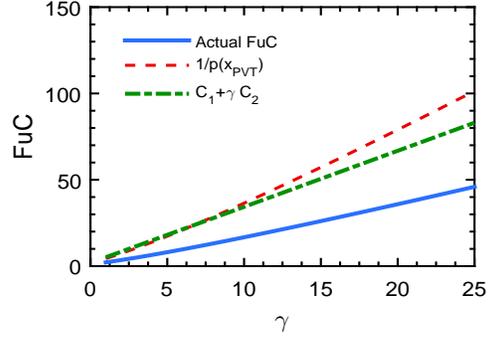}
        \caption{Linear growth of FuC with degree $\gamma$ for \\ $p(x_T) = x_T^\gamma$ and $r(x_T) = 4-x_T$}
        \label{fig:decreasing2}
    \end{subfigure}
    \caption{Behavior of \textit{Fragility under Competition (FuC)} under decreasing rate of return functions: In Figure~\ref{fig:decreasing1} we plot the FuC, the trivial upper bound $\frac{1}{p(x_\mathtt{PVT})}$ and the exponential upper bound in Corollary~\ref{corollary:FuCdecreasing} in a game with a decreasing rate of return function ($r(x_T) = 1.21-0.2x_T$). The risk parameters of the players are $\alpha = 0.88$ and $k = 2.25$. When the failure probability function is $p(x_T)=x^\gamma_T$, FuC grows exponentially with $\gamma$, as shown by a linear growth in the logarithmic scale. In Figure~\ref{fig:decreasing2}, we consider a rate of return function $r(x_T) = 4-x_T$ for which $x^*_r > 1$. As Theorem \ref{theorem:fuclowerbound} suggests, the FuC has an upper bound of the form $C_1 + \gamma C_2$, where $C_1$ and $C_2$ depend on the resource characteristics and the values of $\alpha$ and $k$. The growth of this upper bound and the actual FuC are linear in $\gamma$. The analytical bound becomes tighter than the trivial bound as $\gamma$ increases.}
\end{figure}

\begin{example}
\label{example:fuc_decreasingror}
Consider a Fragile CPR game with $n \geq 2$ homogeneous players each having $\alpha = 0.88$ and $k = 2.25$. We consider $p(x_T) = x^\gamma_T$ for different values of $\gamma$. 

Let the rate of return of the CPR be $r(x_T) = 1.21-0.2x_T$. For this $r(x_T)$, $x^*_r < 1$, and thus Theorem~\ref{theorem:fuclowerbound} indicates that FuC grows exponentially with $\gamma$. In Figure~\ref{fig:decreasing1}, we plot $\frac{p(\bar{y})}{p(x_\mathtt{PVT})}$ as $\gamma$ increases from $1$ to $25$, and compare it with the trivial bound $\frac{1}{p(x_\mathtt{PVT})}$ and the upper bound of $\left(1+\frac{2}{\alpha}\right)^\gamma$ obtained in Corollary~\ref{corollary:FuCdecreasing}. Although the latter bound gets looser as $\gamma$ increases, the actual FuC still grows exponentially in $\gamma$ for values up to $25$. 

Now consider a resource with $r(x_T) = 4-x_T$ or equivalently $\bar{r}(x_T) = (3-x_T)^{0.88}$. In this case, $x^*_r > 1$, and we expect the FuC to grow linearly with $\gamma$. In Figure~\ref{fig:decreasing2}, we plot actual FuC (i.e., $\frac{p(\bar{y})}{p(x_\mathtt{PVT})}$), the trivial bound $\frac{1}{p(x_\mathtt{PVT})}$ and the upper bound obtained in Theorem~\ref{theorem:fuclowerbound}. In this case, as $\gamma$ increases, $\frac{1}{p(x_\mathtt{PVT})}$ gets looser and our analytical upper bound from Theorem~\ref{theorem:fuclowerbound} is tighter. 
\end{example}

\subsubsection{Increasing rate of return} 

As in the analysis for decreasing rate of return, the FuC bound for increasing rate of return depends on the convexity of the failure probability function $p(x_T)$ and is captured by the parameter $\zeta$ defined in Theorem~\ref{theorem:fuclowerbound}. 

\begin{theorem}
Consider a Fragile CPR game $\Gamma$ with concave increasing $\bar{r}(x_T)$, convex strictly increasing $p(x_T)$ and $n \geq 2$ homogeneous players each with sensitivity parameter $\alpha$. Let $\bar{y}$ be the limit of the total PNE investments as $n \to \infty$, and $x_\mathtt{PVT}$ be the optimal investment by a single player in the CPR. Then we have
\begin{enumerate}
\item $\frac{\bar{y}}{x_\mathtt{PVT}} \leq \left(1+\frac{1}{\alpha}\right)$, and  
\item $\mathcal{F}(\Gamma) \leq \left(1+\frac{\zeta}{\alpha}\right)$, 
\end{enumerate}
where $\zeta \triangleq \sup \left\{ \frac{x_Tp'(x_T)}{p(x_T)}, x_T \in (0,1) \right\}$.
\label{theorem:FuCincreasing}
\end{theorem}

As in the proof of Theorem~\ref{theorem:FuCdecreasing}, the proof of the first part relies on the convexity of $p(x_T)$. The second part uses the result of the first part and the definition of $\zeta$. Note that the bound on the ratio of investments is qualitatively similar to the analogous bound obtained for the decreasing rate of return case in that both vary inversely with respect to the sensitivity parameter $\alpha$. The bound on FuC depends on the parameter $\zeta$ in addition to $\alpha$. In particular, the FuC grows linearly with $\zeta$, as in the case with decreasing $r(x_T)$ with $x^*_r > 1$. Note that with $r(x_T)$ increasing in $x_T$, the maximizer of $x_T^\alpha \bar{r}(x_T)$ is unbounded. The following corollary is a direct consequence of the above theorem and is analogous to Corollary~\ref{corollary:FuCdecreasing}.  

\begin{corollary}
Consider a Fragile CPR game $\Gamma$ with homogeneous players, concave increasing $\bar{r}(x_T)$, and polynomial $p(x_T)$ with degree $\gamma$ and all coefficients nonnegative. Then, we have $\mathcal{F}(\Gamma) \leq \left(1+\frac{\gamma}{\alpha}\right)$. 
\label{corollary:FuCincreasing}
\end{corollary}

We obtain the following necessary and sufficient condition under which our analytical bound $1+\frac{\gamma}{\alpha}$ is stronger than the trivial bound $\frac{1}{p(x_\mathtt{PVT})}$.

\begin{proposition}\label{proposition:increasing}
Consider a Fragile CPR game with concave increasing $\bar{r}(x_T)$ and $p(x_T) = x_T^\gamma$. Then $\left(1+\frac{\gamma}{\alpha}\right) \leq \frac{1}{p(x_\mathtt{PVT})}$ if and only if 
$$x_T\bar{r}'(x_T) \leq k\alpha\left(1+\frac{\alpha}{\gamma}\right)$$ 
at $x_T = \left(\frac{\alpha}{\alpha+\gamma}\right)^{\frac{1}{\gamma}}$. In particular, if $\bar{r}(x_T) = ax_T+b$, and $p(x_T) = x_T$, the above condition is equivalent to $a \leq k(1+\alpha)^2$. When $p(x_T) = x^\gamma_T$, $a \leq k\alpha$ is a sufficient condition. 
\end{proposition}

The above result states that our analytical bound from Theorem~\ref{theorem:FuCincreasing} is tighter when the rate of growth of the increasing rate of return function is not too large. In particular, when the rate of return is constant, the above condition is always satisfied. In fact, the bounds obtained in Theorem~\ref{theorem:FuCincreasing} and Corollary~\ref{corollary:FuCincreasing} on $\frac{\bar{y}}{x_\mathtt{PVT}}$ and $\frac{p(\bar{y})}{p(x_\mathtt{PVT})}$ are {\it tight} when the rate of return function is a constant, as we demonstrate in the following example. Subsequently, we also demonstrate through numerical examples that for rate of return functions where the trivial bound $\frac{1}{p(x_\mathtt{PVT})}$ is tighter, the growth of this trivial bound and the actual FuC are also linear with respect to the exponent $\gamma$ of the failure probability function.  

\begin{example}
\label{example:fuc_increasingror}
Consider a Fragile CPR game $\Gamma$ with $n$ homogeneous players each having risk parameters $\alpha$ and $k$. Let $r(x_T) = b+1, b>0$ and $p(x_T)=x_T^\gamma$, $x_T \in [0,1], \gamma \geq 1$. 
The expected utility~\eqref{eq:expectedutility} of player $i$ is, 
$$\mathbb{E}(u_{i}) = x_i^\alpha[b^\alpha(1-x_T^\gamma)-kx_T^\gamma] = x_i^\alpha[b^\alpha-(b^\alpha+k)x_T^\gamma].$$
Let the total investment at the PNE be denoted as $x^*_{Tn}$, parametrized by $n$. Since players have homogeneous risk preferences, the investment by each individual player must be $\frac{x^*_{Tn}}{n}$. The total investment at the PNE satisfies the first order condition of optimality in~\eqref{eq:homogeneous_foc_PNE}, i.e., 
\begin{align*}
& 0 = \frac{x^*_{Tn}}{n}f'(x^*_{Tn})+\alpha f(x^*_{Tn}) = \left[ - \frac{x^*_{Tn}}{n}(b^\alpha+k)\gamma (x^*_{Tn})^{\gamma-1} + \alpha(b^\alpha - (b^\alpha+k)(x^*_{Tn})^\gamma) \right]
\\ \implies & (x^*_{Tn})^\gamma \left(\frac{\gamma}{n}+\alpha\right) = \frac{\alpha b^\alpha}{b^\alpha+k}.
\end{align*}
Let the optimal investment by a single player be denoted as $x_\mathtt{PVT}$. Then by setting $n=1$ in the above equation, we have
$$ (x_\mathtt{PVT})^\gamma (\gamma+\alpha) = \frac{\alpha b^\alpha}{b^\alpha+k}. $$
Thus we have
$$ \mathcal{F}(\Gamma) = \frac{(x^*_{Tn})^\gamma}{(x_\mathtt{PVT})^\gamma} =\frac{\gamma+\alpha}{\alpha+\frac{\gamma}{n}}, $$
and when $n \to \infty$, the FuC converges to $\left(1+\frac{\gamma}{\alpha}\right)$. Note that when $p(x_T) = x_T$ (i.e., $\gamma = 1$), we have
$$ \frac{(x^*_{Tn})^\gamma}{(x_\mathtt{PVT})^\gamma} =\frac{1+\alpha}{\alpha+\frac{1}{n}} \to 1+\frac{1}{\alpha} \text{   as  } n\to\infty. $$
\end{example}

The above example demonstrates that when the rate of return is independent of the total investment by the players, i.e., $r(x_T) = b+1, b>0$ and $p(x_T) = x_T$, the ratio of total investment at the PNE and the optimal investment by a single player approaches the upper bound obtained in Theorem~\ref{theorem:FuCincreasing} as $n \to \infty$. Furthermore, for $p(x_T) = x_T^\gamma$, the FuC also approaches the upper bound obtained in Corollary~\ref{corollary:FuCincreasing}. In the following example, we compare the growth of FuC, the bound obtained in Theorem~\ref{theorem:FuCincreasing}, and $\frac{1}{p(x_\mathtt{PVT})}$ as $\gamma$ increases.  

\begin{figure}[t!]
    \begin{subfigure}[t]{0.5\textwidth}
	\includegraphics[width=7.8cm,height=5cm]{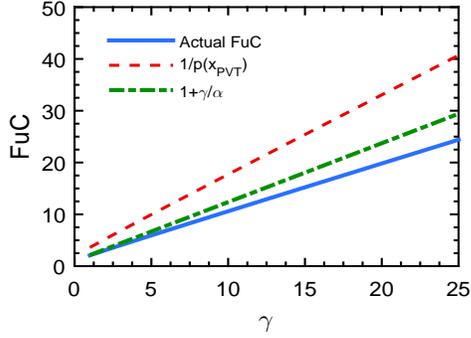}
        \caption{Comparison of trivial and analytical bounds for \\ $p(x_T) = x_T^\gamma$ and $r(x_T) = x_T+4$.}
        \label{fig:FuCIncreasing1}
    \end{subfigure}%
    ~ 
    \begin{subfigure}[t]{0.5\textwidth}
        \includegraphics[width=7.8cm,height=5cm]{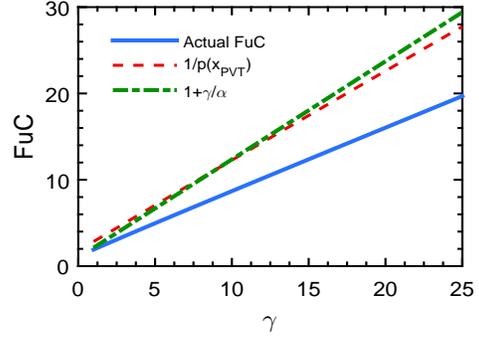}
        \caption{Comparison of trivial and analytical bounds for \\ $p(x_T) = x_T^\gamma$ and $r(x_T) = 4x_T+4$.}
        \label{fig:FuCIncreasing2}
    \end{subfigure}
    \caption{Behavior of \textit{Fragility under Competition (FuC)} under increasing rate of return functions: In Figure \ref{fig:FuCIncreasing1}, we plot the actual FuC, the trivial bound $\frac{1}{p(x_\mathtt{PVT})}$ and the analytical bound $1+\frac{\gamma}{\alpha}$ obtained in Theorem~\ref{theorem:FuCincreasing} for a resource with $r(x_T)= x_T+4$, and players having $k=2.25$ and $\alpha = 0.88$. Our analytical bound is tighter than the trivial bound for all values of $\gamma$ between $1$ to $25$, in accordance with the predictions in Proposition~\ref{proposition:increasing}. For $r(x_T)= 4x_T+4$, the rate of growth of $r(x_T)$ is higher. As a result the trivial bound becomes tighter as the values of $\gamma$ increase, as shown in Figure \ref{fig:FuCIncreasing2}. In both plots, the growth of actual FuC and the trivial bound are linear in $\gamma$. }
\end{figure}

\begin{example}
\label{example:increasingr_trivialbound}
Consider a Fragile CPR game with homogeneous players. Let $k=2.25$ and $\alpha = 0.88$. We consider $p(x_T) = x^\gamma_T$. As $\gamma$ grows from $1$ to $25$, we plot $\frac{p(\bar{y})}{p(x_\mathtt{PVT})}$ (actual FuC when the number of players $n\to\infty$), $1+\frac{\gamma}{\alpha}$ and $\frac{1}{p(x_\mathtt{PVT})}$ for two different rate of return functions; Figure~\ref{fig:FuCIncreasing1} shows our plot for $r(x_T) = x_T+4$ and Figure \ref{fig:FuCIncreasing2} shows our plot for $r(x_T) = 4x_T+4$. 

As we see from Figure~\ref{fig:FuCIncreasing1}, for $r(x_T) = x_T+4$, the upper bound $1+\frac{\gamma}{\alpha}$ is a stronger bound than the trivial bound $\frac{1}{p(x_\mathtt{PVT})}$ for all values of $\gamma$ from $1$ to $25$. On the other hand, Figure \ref{fig:FuCIncreasing2} shows that for $r(x_T) = 4x_T+4$ (which has a higher rate of increase than $r(x_T) = x_T+4$) the trivial bound $\frac{1}{p(x_\mathtt{PVT})}$ becomes stronger as $\gamma$ increases, in particular for $\gamma \geq 10$. Nonetheless, $\frac{1}{p(x_\mathtt{PVT})}$ grows linearly with $\gamma$ even when it is the stronger of the two bounds, as does the actual FuC. 
\end{example}

\subsubsection{Discussion}
The results obtained above offer several interesting insights. First we showed that total investment in the CPR at the PNE increases with the number of players competing for the resource, illustrating the tragedy of the commons phenomenon. For both increasing and decreasing rate of return functions, the total investment by any number of players is bounded by a certain multiple of the optimal investment by a single player, and these multiplicative factors are independent of the characteristics of the CPR (the rate of return and failure probability functions, provided they satisfy Assumption~\ref{assumption:CDCI}). Furthermore, the bounds are independent of the loss aversion index $k$, and inversely proportional to the sensitivity parameter $\alpha$, which characterizes the risk aversion/seeking behavior under prospect theory. As players become more prospect-theoretic, i.e., increasingly risk averse in gains and risk seeking in losses, captured by $\alpha$ decreasing below $1$, the bounds on the ratio of investments (and FuC) increase. In other words, the bounds are larger when accounting for prospect-theoretic risk preferences of the players compared to when players are risk neutral or classical expected utility maximizers. 

As opposed to the above bounds on the ratio of investments, the bounds on FuC depend on the resource characteristics in addition to the prospect-theoretic preferences of players. Specifically when $p(x_T) = x_T^\gamma$, the upper bound on FuC is linearly increasing in $\gamma$ when $r(x_T)$ is increasing in total investment. For decreasing $r(x_T)$, the bounds depend on the value of the maximizer of $x_T^\alpha \bar{r}(x_T)$, which captures the investment in the CPR by a single user in the absence of resource failure. When the maximizer of $x_T^\alpha \bar{r}(x_T)$ is larger than $1$, the FuC has a different upper bound which is also linear in $\gamma$. On the other hand, when the maximizer of $x_T^\alpha \bar{r}(x_T)$ is smaller than $1$, we obtained an asymptotic lower bound which grows exponentially in $\gamma$ as the number of players grows to infinity. These different rates of growth are, in fact, observed in the Fragile CPR games considered in Example~\ref{example:increasingr_trivialbound} and Example~\ref{example:fuc_decreasingror}, respectively. The explanation for the difference is that when $r(x_T)$ is increasing, a player investing in isolation is already aggressive with her investment, thereby limiting the extra investment possible in a competitive environment. In contrast, when an isolated player is conservative with her investment even in the absence of resource failure for certain decreasing $r(x_T)$, a relatively large portion of unused investment is expended in a competitive environment (at the PNE). 

\subsection{Fragility at social optimum}

So far, we have compared the fragility of the CPR when there are a large number of homogeneous players competing for the resource to the fragility when the resource is used by a single player. A related question of interest is to quantify how large the failure probability gets under selfish decision making (PNE) compared to the case where investment decisions are made by a social planner (centralized optimization), for a given number of players. The social planner maximizes a social welfare function, which is commonly defined as the sum of utilities of all players~\cite{nisan2007algorithmic}. We denote the social welfare function as $\Psi(x_1,x_2,\ldots,x_n) = \sum_{i \in \mathcal{N}} \mathbb{E}(u_{i})$. In the following result, we show that when players have homogeneous risk preferences, the total investment at the social optimum is identical to the optimal investment by a single player in the resource. 

\begin{proposition} 
\label{proposition:socialwelfarealpha}
Consider a Fragile CPR game with homogeneous players. The total investment at the social optimum is independent of the number of players and is equal to the optimal investment by a single player.
\end{proposition}

The proof is presented in~\ref{appendix:fucproof} and follows from the first order necessary condition of optimality and homogeneous risk preferences of the players. The above result, in conjunction with our bounds in Theorems~\ref{theorem:FuCdecreasing} and~\ref{theorem:FuCincreasing}, offers the following interpretation. The investments prescribed by a central planner, although not chosen to directly minimize fragility, nevertheless result in a more robust resource as compared to the case under selfish decentralized decision making. Furthermore, for certain rate of return functions, a social planner becomes increasingly beneficial (in terms of reducing fragility) as $\alpha$ decreases (i.e., as players become less sensitive to large gains and losses). In other words, the benefit of a social planner becomes even larger under prospect-theoretic players ($\alpha < 1$) in comparison to a setting with risk neutral players ($\alpha = 1$).

Proposition~\ref{proposition:socialwelfarealpha} also has consequences for the so-called \textit{Price of Anarchy} of the game, which is a well established metric that quantifies loss of welfare at the PNE compared to the social optimum~\cite{nisan2007algorithmic}. This metric is defined as the ratio of the highest achievable social welfare of the game (i.e., at the social optimum) and the minimum social welfare obtained at any PNE. Thus it captures the inefficiency of the PNE in terms of the loss of utilities of the players. However it gives little indication of the failure probability of the resource at equilibrium, which is a form of externality caused by selfish decision making by the players. In a CPR setting, fragility of the resource is of interest to the social planner in addition to the welfare derived from the utilities of the players. Our FuC metric bridges this gap by studying the effect of selfish decision making by the players and their risk preferences on the fragility of the resource. For completeness, we include a brief discussion of price of anarchy in~\ref{appendix:poa} and show that, in general, it is unbounded in the number of players. The result follows from our observation in Proposition~\ref{proposition:boundedpneXT}.

\section{Fragility Under Heterogeneity}
\label{section:fuh}
In the previous section, we isolated the effect of the number of players on the fragility of the CPR, while keeping all risk attitudes homogeneous. We devote this section to studying the effects of heterogeneity in risk attitudes, while keeping the number of players fixed. Specifically, we investigate how a heterogeneous society compares to its homogeneous counterpart, vis-a-vis the fragility of the CPR. We study heterogeneity in the loss aversion indices of players first, followed by heterogeneity in the sensitivity parameters. The proofs of all results of this section are presented in~\ref{appendix:fuhproofs}. 

\subsection{Heterogeneity in loss aversion indices}
\label{subsection:lossaversionhet}

In this subsection, we consider a Fragile CPR game with $n$ players, where the sensitivity parameter $\alpha$ is identical across players, i.e., $\alpha_1 = \alpha_2 = \ldots = \alpha_n = \alpha$. We refer to such games as $\alpha$-uniform Fragile CPR games. Without loss of generality, let players be labeled so that $k_1 \leq k_2 \leq \ldots \leq k_n$. We will refer to a player with index of loss aversion $k_i$ as player $i$ for convenience. We start by examining the effect of loss aversion on the investment of a player at the PNE. 

\begin{proposition}\label{proposition:lossaversexi}
Consider an $\alpha$-uniform Fragile CPR game, and let two players $i$ and $j$ have $k_i \leq k_j$. Then the investment of player $i$ at the PNE is at least as high as the investment of player $j$. In particular, if $x^*_i = 0$, then a player $j$ with $k_j \geq k_i$ has $x^*_j = 0$.
\end{proposition}

The proof of this result follows from examining how the index of loss aversion affects the effective rate of return function and its derivative. In fact the monotonicity of the failure probability function is sufficient to establish the above proposition; we do not need the monotonicity and concavity assumptions on the rate of return functions. The result shows that the support of the PNE in an $\alpha$-uniform game consists of a set of players with smallest loss aversion indices (recall that the support of a PNE is defined as the set of players with nonzero investment). In other words, sufficiently loss averse players get driven out of the CPR by gain seeking players, and instead prefer to invest entirely in the safe resource. 

Our goal is to contrast the fragility of the CPR in $\alpha$-uniform games when players have homogeneous loss aversion indices to the case where they differ in their attitudes towards loss. For a fair comparison, we take the mean loss aversion index in the heterogeneous society to be equal to the loss aversion index in the homogeneous society. Specifically, consider an $\alpha$-uniform Fragile CPR game $\Gamma_M$ with $n$ homogeneous players, where each player has loss aversion index $k = k_M$. Let $\mathcal{K}$ be defined as
\begin{equation}\label{eq:meanpreservingk}
\mathcal{K} = \left\{ \{k_1,k_2,\ldots,k_n\} \in \mathbb{R}^n_{\geq 0} | k_1 \leq k_2 \leq \ldots \leq k_n, \frac{1}{n}\sum_{i=1}^n k_i = k_M \right\}, 
\end{equation}
i.e., $\mathcal{K}$ is the set of all feasible $n$-dimensional vectors of loss aversion indices with mean $k_M$. Define $\boldsymbol{\Gamma}$ to be the set of $\alpha$-uniform Fragile CPR games with the vector of loss aversion indices of players drawn from the set $\mathcal{K}$, and all other parameters the same as in the original game $\Gamma_M$. Clearly $\Gamma_M \in \boldsymbol{\Gamma}$. In the following result, we show that $\Gamma_M$ has the smallest fragility among all games in $\boldsymbol{\Gamma}$. To avoid trivial games with nonpositive investment in the CPR at the PNE, we assume that $k_M$ is chosen such that for a player $i$ with $k_i = k_M$, $f_{i}(x_T) > 0$ for some $x_T \in [0,1]$.  

\begin{theorem}
Let $\boldsymbol{\Gamma}$ be a class of $\alpha$-uniform Fragile CPR games with loss aversion indices of players having mean $k_M$, i.e., they are drawn from the set $\mathcal{K}$ defined in~\eqref{eq:meanpreservingk}. Let $\Gamma_M \in \boldsymbol{\Gamma}$ be the game where players have homogeneous loss aversion indices. Then the fragility of $\Gamma_M$ is smallest across all games in $\boldsymbol{\Gamma}$. 
\label{theorem:fuh}
\end{theorem}

The proof relies on the convexity of the PNE investment of a player (captured by the function $g_i(x_T)$ defined in \eqref{eq:focbestresponse}) with respect to her loss aversion index. We conclude our study of the effect of the loss aversion index on the fragility of the CPR by proving the following intuitive result. Its proof relies on the monotonicity of the best response established in Lemma~\ref{lemma:garg}. 

\begin{proposition}
Consider two Fragile CPR games $\Gamma_1$ and $\Gamma_2$ with identical resource characteristics. Let both games have $n$ homogeneous players such that players have identical $\alpha$ across both games, while the players in $\Gamma_2$ are more loss averse, i.e., have larger $k$, than their counterparts in $\Gamma_1$. Then the resource is less fragile at the PNE of $\Gamma_2$ compared to the PNE of $\Gamma_1$. 
\label{proposition:alphauniformfragility}
\end{proposition}

\subsection{Heterogeneity in sensitivity parameter $\alpha$}

The proofs of our results in Subsection~\ref{subsection:lossaversionhet} relied on the property that the effective rate of return function $f$ and its derivative $f'$ were both decreasing in the loss aversion index of the player. As a result, we could show that the equilibrium investment in an $\alpha$-uniform game is also decreasing in the loss aversion index of a player. This observation was key to the proof of Theorem~\ref{theorem:fuh}, where we showed that homogeneity in loss aversion results in less fragility compared to the case where players have heterogeneous loss aversion indices (provided the vectors of loss aversion indices have identical mean). However, the functions $f$ and $f'$ do not exhibit such monotonicity in the sensitivity parameter $\alpha$ (as can be observed from their expressions in equations~\eqref{eq:expectedutility} and~\eqref{eq:effectiverormonotonicity}). Therefore, the counterparts of the result in Theorem~\ref{theorem:fuh} do not hold, as illustrated by the following counterexample. In addition, the counterpart of the result in Proposition~\ref{proposition:alphauniformfragility} also does not hold, as we show in Example~\ref{example:fragilityalpha}.

\begin{example} In this example, we show that when players have identical loss aversion indices, heterogeneity in $\alpha_i$ can result in either a decrease or an increase in fragility of the resource compared to the fragility under homogeneous players.

Consider a Fragile CPR game with $n=3$ players. Let the rate of return function be $r(x_T) = 5-x_T$ and the failure probability function be $p(x_T) = x_T$. Let $k_i = 1, i \in \{1,2,3\}$, i.e., players have identical loss aversion indices. In Table~\ref{table:heteroalphadecreasing1}, we show that the fragility increases as players become heterogeneous in their sensitivity parameter. On the other hand, under a different rate of return function $r(x_T) = 1.55-0.5x_T$ (with the same $p(x_T)$ and loss aversion indices as above), fragility is found to decrease with heterogeneity in $\alpha_i$. 

Similarly, when the rate of return is increasing, we give two examples in Table~\ref{table:heteroalphaincreasing1} which show both increase and decrease in fragility under heterogeneity for different $r(x_T)$. Both examples have $p(x_T) = x_T$ and $k_i = 1, i \in \{1,2,3\}$.

\begin{table}
	\begin{subtable}{1\linewidth}
	\centering
		{\begin{tabular}{| c | c | c |}
			\hline
			Risk preference   & Fragility for $r(x_T) = 5-x_T$ & Fragility for $r(x_T) = 1.55-0.5x_T$  \\ \hline
			$\alpha_1 = 0.5, \alpha_2 = 0.5, \alpha_3 = 0.5$ &  $0.3846$ & 0.2233 \\ \hline
			$\alpha_1 = 0.3, \alpha_2 = 0.3, \alpha_3 = 0.9$ &  $0.4018$ & 0.2083 \\ \hline
		\end{tabular}}
		\caption{Fragility under heterogeneity in $\alpha$ for decreasing $r(x_T)$}
		\label{table:heteroalphadecreasing1}
	\end{subtable} 
	
	\vspace{4mm}
	
	\begin{subtable}{1\linewidth}
	\centering
	{
	\begin{tabular}{| c | c | c|}
	\hline
	Risk preference   & Fragility for $r(x_T) = 3+x_T$ & Fragility for $r(x_T) = 1.05+0.9x_T$ \\ \hline
	$\alpha_1 = 0.5, \alpha_2 = 0.5, \alpha_3 = 0.5$ &  $0.3758$ & 0.2481 \\ \hline
	$\alpha_1 = 0.3, \alpha_2 = 0.3, \alpha_3 = 0.9$ &  $0.4035$ & 0.2014 \\ \hline
	\end{tabular}}
	\caption{Fragility under heterogeneity in $\alpha$ for increasing $r(x_T)$}
	\label{table:heteroalphaincreasing1}
	\end{subtable}%
\caption{Fragility under heterogeneity in the sensitivity parameter $\alpha$: In Table~\ref{table:heteroalphadecreasing1}, we give two numerical examples of CPRs with decreasing rates of return for which the fragility could either increase (example in middle column) or decrease (example in last column) with heterogeneity in $\alpha$ when compared to a game where $\alpha$ is homogeneous among players and is the mean of the heterogeneous values. We give two analogous examples for increasing rates of return in Table~\ref{table:heteroalphaincreasing1}. In all of the examples in the table, $p(x_T) = x_T$ and the indices of loss aversion of the players are chosen to be $1$.}
\end{table}
\end{example}

\begin{figure}[t]
    \centering
    \begin{subfigure}[t]{0.49\textwidth}
	\centerline{\includegraphics[width=6cm,height=5cm]{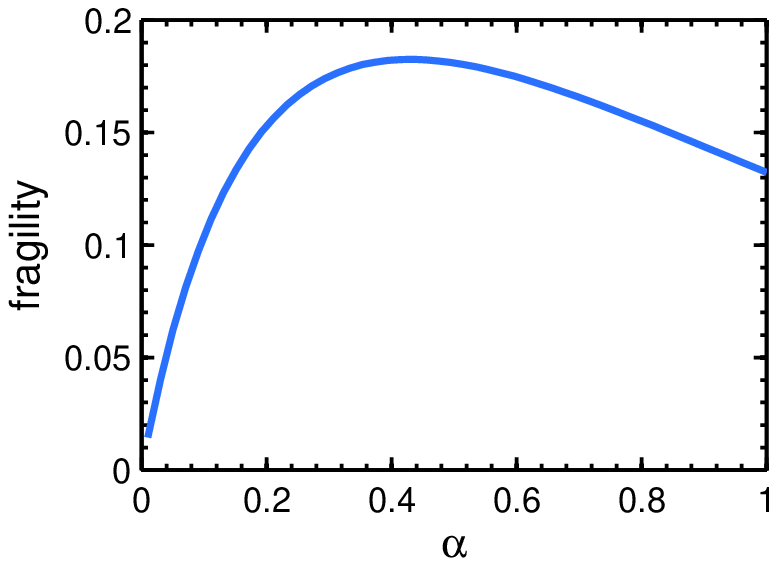}}
	\caption{Variation of fragility with $\alpha$ for $r(x_T) = 1.25-0.2x_T$}
	\label{fig:fuhCSdecreasing}
    \end{subfigure}%
    ~ 
    \begin{subfigure}[t]{0.49\textwidth}
	\centerline{\includegraphics[width=6cm,height=5cm]{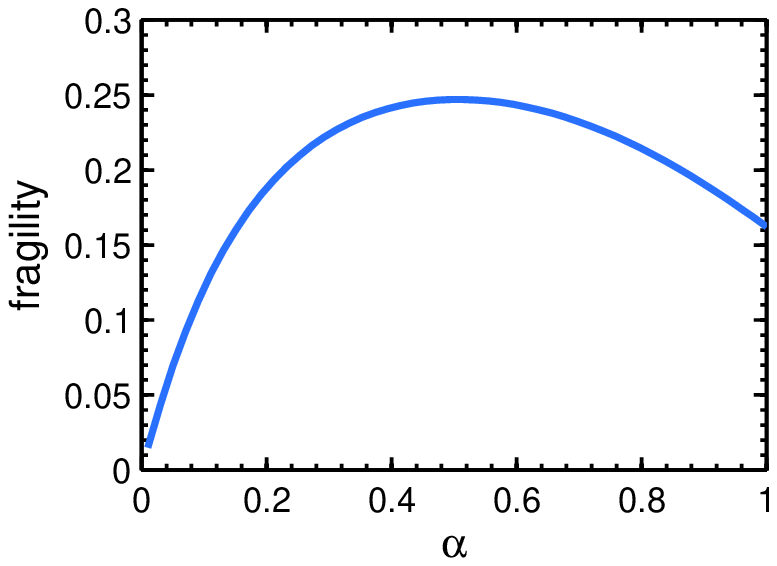}}
	\caption{Variation of fragility with $\alpha$ for  $r(x_T) = 1.1+0.8x_T$}
	\label{fig:fuhCSincreasing}
    \end{subfigure}
    \caption{Non-monotonicity of fragility with respect to the sensitivity parameter: In Figure~\ref{fig:fuhCSdecreasing} we plot the failure probability at the PNE of a game with three homogeneous players and a decreasing rate of return $r(x_T) = 1.25-0.2x_T$, as defined in Example~\ref{example:fragilityalpha}. As we increase the sensitivity parameter $\alpha$ uniformly among the players, the fragility first increases and then decreases. In Figure~\ref{fig:fuhCSincreasing}, we illustrate similar behavior in a game with increasing rate of return $r(x_T) = 1.1+0.8x_T$. In both examples, the failure probability function is given by $p(x_T) = x_T$ and the loss aversion indices of the players are $1$.}
\end{figure}

\begin{example}\label{example:fragilityalpha}
In Proposition~\ref{proposition:alphauniformfragility} we showed that in a game with homogeneous players and fixed $\alpha$, an increase in the (common) loss aversion index results in a decrease in the failure probability of the resource at the PNE. Here we give two examples showing that change in fragility need not be monotonic in $\alpha$. 

First we give an example where the rate of return is decreasing in total investment. Consider a game with $n=3$ players and $k_i = 1, i \in \{1,2,3\}$. Let the resource have $r(x_T) = 1.25-0.2x_T$ and $p(x_T)=x_T$. Figure~\ref{fig:fuhCSdecreasing} shows the change in fragility as we vary $\alpha$ from $0$ to $1$ uniformly for all players. With an increase in $\alpha$ (i.e., as players move closer to risk neutral behavior), fragility increases first and then decreases. Similar behavior is observed with increasing rate of return functions, when $r(x_T) = 1.1+0.8x_T$, $p(x_T)=x_T$ and $k_i = 1, i \in \{1,2,3\}$, as shown in Figure~\ref{fig:fuhCSincreasing}. 

It is worth noting that there are also examples in which fragility can exhibit monotonic (increasing or decreasing in $\alpha$) behavior depending on the rate of return function, number of players, etc. 
\end{example}

\section{Conclusion}
\label{section:conclusion}

In this paper, we studied a common-pool resource (CPR) game with probabilistic resource failure due to investment in the resource. Players exhibit potentially heterogeneous risk attitudes, captured by the value function from prospect theory. We proved the existence and uniqueness of a pure Nash equilibrium (PNE), and showed that this class of games falls within the class of best response potential games, in which sequential best response dynamics converge to the PNE. 

Competition among multiple players in the game implies greater total investment (and probability of failure) in the CPR at the PNE as compared to the optimal investment by a single player. In order to characterize the effect of greater competition on the fragility of the CPR, we introduced the metric of Fragility under Competition (FuC), and obtained upper bounds on FuC for both increasing and decreasing rate of return functions. A high FuC potentially results when i) the players become increasingly risk averse in gains and risk seeking in losses (i.e., deviate from risk neutral and classical expected utility maximization behavior), and ii) the resource exhibits a sharp transition from a relatively safe state to a fragile state (e.g., tipping point behavior). Our analysis shows that for certain classes of resources, the investments prescribed by a social welfare maximizer also serve to reduce the fragility of the resource (as compared to the investments at the PNE); furthermore, this reduction in fragility becomes larger as players deviate more from risk neutrality (as captured by the prospect-theoretic sensitivity parameter $\alpha$).

We also examined the role of heterogeneity in sensitivity and attitudes towards loss aversion on the fragility of the CPR. A society where users have heterogeneous loss aversion indices invests more in the CPR compared to a society with identical loss aversion indices, resulting in greater fragility. However, heterogeneity in the sensitivity parameter $\alpha$ can result in either higher or lower fragility depending on the parameters of the game. 

There are several avenues for further exploration in the setting considered in this paper. 

\textit{Probability Weighting:} Under prospect theory (and cumulative prospect theory~\cite{tversky1992advances}), users maximize the expected value function with subjective probabilities, rather than actual probabilities. The resulting weighting function overweights small probabilities and underweights moderate to large probabilities. As a result, the shape of the function is concave in the low probability region, and is convex for moderate to high probabilities. This presents challenges for analysis, and thus developing techniques to study the effect of probability weighting is a natural and important next step for research. 

\textit{Bounded Endowment:} In our setting, the initial endowment of each player is sufficiently large so that it does not explicitly constrain her investment in the CPR. The implicit constraint is the investment at which the resource fails with certainty. On the other hand, there can be settings where individual endowment is not large enough to have a significant impact on the failure probability of the resource. Studying the effects of bounded endowments on players' strategies will likely uncover additional interesting phenomenon. For instance, in resources with increasing rate of return, players with small endowments may not invest unless a sufficiently large number of other players have invested. This leads to the violation of our key result in Lemma~\ref{lemma:interval} and the monotonicity of the best response in the investment of other players. As a result, there could be multiple pure Nash equilibria. 

Risk preferences of the players could be correlated with the endowment bounds, and heterogeneity in the endowments could have important implications for the fragility of the resource. Similarly, if the bounds of the players are normalized and decrease so that individual endowments tend to $0$ as the number of players $n \to \infty$, the resulting game approaches the ``nonatomic congestion game" setting~\cite{roughgarden2004bounding}, a model often used to study selfish routing in networks. These observations further motivate a future study of bounded endowments in fragile CPR games.  

\textit{Mechanism Design:} While we studied the effects of competition and heterogeneous risk preferences on the fragility of the CPR, an equally interesting question is how to design mechanisms for players with private prospect-theoretic risk attitudes in order to obtain socially desirable outcomes (such as ensuring low failure probability of the resource or reducing risky behavior by investors) at the PNE. 

\textit{Aggregative Games:} Finally our work motivates further exploration of prospect-theoretic risk attitudes in a broad class of games with similar mathematical structure as ours, where the utility of a player is a function of the sum of strategies of all other players. Such games are often referred to as {\it aggregative games} and several well-studied classes of problems, such as private provisioning of public goods~\cite{bergstrom1986private}, rent seeking contests~\cite{nitzan1994modelling} and a class of interdependent security games~\cite{grossklags2008security} fall into this category. Our proof techniques and results might provide useful insights into the behavior of heterogeneous prospect-theoretic agents in the above settings. In security games with multiple stakeholders, our FuC metric might have applicability as a measure of inefficiency of equilibria to quantify potential increase in the fragility of the system due to distributed decision making.  

\section*{Acknowledgements}
We thank Prof. Dawn Parker (University of Waterloo) and Prof. Siddhartha Banerjee (Cornell University) for helpful discussions. We also thank the anonymous reviewers for their suggestions which improved the paper. 

\bibliographystyle{plainnat}
{\footnotesize \bibliography{refs,refs2}}

\appendix
\section{Proofs pertaining to the existence and uniqueness of the PNE}
\label{appendix:existenceproof}

\noindent \textbf{Proof of Lemma~\ref{lemma:interval}.}
We consider increasing and decreasing rate of return functions separately. We will use properties of the first and second derivatives of the effective rate of return~\eqref{eq:expectedutility}, given by 
\begin{align}
\frac{d f(x_T)}{d x_T} &= \bar{r}'(x_T)(1-p(x_T)) - (\bar{r}(x_T)+k) p'(x_T), \label{eq:effectiverormonotonicity}
\\ \frac{d^2 f(x_T)}{d x_T^2} & = \bar{r}''(x_T)(1-p(x_T)) - 2\bar{r}'(x_T)p'(x_T) - (\bar{r}(x_T)+k) p''(x_T). \label{eq:effectiverorconcavity} 
\end{align}

Recall from Remark~\ref{remark:concavityofrbar} that $\bar{r}(x_T) = (r(x_T)-1)^\alpha$ is concave increasing (decreasing) if $r(x_T)$ is concave increasing (decreasing). 

\textbf{Case 1:} {\textit{$\bar{r}(x_T)$ is decreasing}.}

In this case, it is easy to see from~\eqref{eq:effectiverormonotonicity} that the effective rate of return $f$ is strictly decreasing for $x_T \in [0,1)$, since $p'(x_T)>0$. If $f(0) \leq 0$, then the only best response of the player is to invest entirely in the safe resource, so in this case $\bar{y}=0$, and the interval $\mathcal{I}$ is not defined. 

Now consider effective rate of return functions for which $f(0) > 0$. Since $f(1) = -kp(1) = -k \leq 0$, $\exists \bar{y} \in (0,1]$ such that $f(\bar{y})=0$. Thus when $y \geq \bar{y}$, we have $B(y) = \{0\}$, as any positive investment would result in nonpositive utility for the player. The best response correspondence is single-valued in this case. 

On the other hand when $y<\bar{y}$, $f(x+y)>0$ for $0<x<\bar{y}-y$. Since there exists $x$ for which $\mathbb{E}(u)=xf(x+y)>0$, every $b(y) \in B(y)$ is strictly positive in the range $y \in [0,\bar{y})$. Thus, $0 \notin B(y)$ if and only if $y \in [0,\bar{y})$. In addition, $x_T=(b(y)+y) \in (0,\bar{y})$ for all $b(y)\in B(y)$ since expected utility must be positive at the nonzero best response. We denote the interval $(0,\bar{y})$ as $\mathcal{I}$ (e.g., see Figure~\ref{fig:lemmainterval1}).
\label{case:lemmaintervalcasedecreasing}

\smallskip

\textbf{Case 2:} {\textit{$\bar{r}(x_T)$ is increasing}.}

Here $f(x_T)$ is concave in $x_T$, i.e., $f''(x_T) \leq 0$, as we see from~\eqref{eq:effectiverorconcavity} (recall that $\bar{r}(x_T) > 0$ from Assumption~\ref{assumption:CDCI}). 

If $f(x_T) \leq 0, \forall x_T\in[0,1]$, then investing $0$ is a best response for the player irrespective of the total investment by other players. Therefore, we define $\bar{y}=0$, and $\mathcal{I}$ is undefined here. Thus suppose $f(x_T)>0$ for some $x_T \in [0,1]$, and define $\hat{x}_T := \inf \{x \in [0,1] | f(x)>0\}$. 

Note that if $\hat{x}_T >0$, then $f'(\hat{x}_T)>0$ by the definition of $\hat{x}_T$. Now there are two possible cases arising from whether or not $f'(\hat{x}_T)$ is negative, and we analyze them separately. 
\begin{itemize}
\item Suppose $f'(\hat{x}_T) \geq 0$. Since $\bar{r}$ is concave, positive, increasing and continuous in $[0,1]$, and since $p(1) = 1$, we see from~\eqref{eq:effectiverormonotonicity} that for $\delta$ sufficiently close to $1$, $f'(\delta) < 0$. Since $f$ is concave in $[\hat{x},1), \exists z \in [\hat{x}_T,\delta]$ such that $f'(z)=0$. Define $\hat{z} := \inf\{z\in(\hat{x}_T,1)|f'(z)<0\}$. In addition, since $f(1) \leq 0$, $\exists \bar{y} \in (\hat{z},1]$ where $f(\bar{y})=0$ (e.g., see Figure~\ref{fig:lemmainterval2}). 
\item Otherwise suppose $\hat{x}_T=0$ and $f'(\hat{x}_T) < 0$. The argument here is similar to the previous case with decreasing rate of return. We define $\hat{z}=0$ and it is easy to see that $\exists \bar{y} \in (0,1]$ with $f(\bar{y})=0$. 
\end{itemize}
When the total investment $y$ of all other players satisfies $y \geq \bar{y}$, then investing zero in the CPR is the only best response, i.e., $B(y) = \{0\}$ (single-valued) as before. Otherwise $b(y) >0$ for all $b(y) \in B(y)$, using a similar argument as in the previous case. Here, we must have $f(b(y)+y)>0$ and $f'(b(y)+y)<0$, as argued after~\eqref{eq:pnexi}. We define $\mathcal{I}$ as the region of interest in which $f'(x_T)<0$ and $f(x_T)>0$, i.e., $\mathcal{I} = (\hat{z},\bar{y}) \subset [0,1]$. Thus, whenever $b(y)>0$, $y+b(y) \in \mathcal{I}$. 
\qed

\vspace{3mm}

\noindent \textbf{Proof of Lemma~\ref{lemma:CDuniquebr}.}
We use Lemma~\ref{lemma:interval} to show that the expected utility has a single maximum in the interval $\mathcal{I}$. If $f(x_T) \leq 0, \forall x_T \in [0,1]$, we have $\bar{y} = 0$ for both decreasing and increasing rate of return functions. In this case, we define the best response to be zero and single-valued, i.e., $B(y)=\{0\}, \forall y \in \bar{S}$, even if we have $f(x_T)=0$ for some $x_T>0$ in the increasing rate of return case. On the other hand when $0 \leq y < \bar{y}$, we have nonzero best responses. 

\smallskip
\textbf{Case 1:}  {\textit{$\bar{r}(x_T)$ is decreasing}.}

When any best response $b(y) \in B(y)$ is nonzero, it is a solution to $\frac{\partial \mathbb{E}(u)}{\partial x}=0$, which is given by
\begin{align}
\frac{\partial \mathbb{E}(u)}{\partial x} &= [x^\alpha \bar{r}'(x+y)+\alpha x^{\alpha-1}\bar{r}(x+y)](1-p(x+y)) - x^\alpha \bar{r}(x+y)p'(x+y) \nonumber
\\ & \qquad -k[\alpha x^{\alpha-1} p(x+y)+x^\alpha p'(x+y)] = 0.
\label{eq:prospect_firstderivative}
\end{align}
It is easy to see that for $\bar{r}$ and $p$ satisfying the conditions in Assumption~\ref{assumption:CDCI}, the last two terms in~\eqref{eq:prospect_firstderivative} are strictly negative. Thus, any solution $x=b(y) > 0$ must have $x^{\alpha-1}[x \bar{r}'(x+y)+\alpha \bar{r}(x+y)]>0$. However, in the region where $x^{\alpha-1}[x\bar{r}'(x+y)+\alpha \bar{r}(x+y)]>0$, we have
\begin{align}
\frac{\partial^2 \mathbb{E}(u)}{\partial x^2} &= [x^\alpha \bar{r}''(x+y)+2 \alpha x^{\alpha-1} \bar{r}'(x+y)](1-p(x+y)) \nonumber
\\  & \qquad -2[x^{\alpha}\bar{r}'(x+y)+\alpha x^{\alpha-1} \bar{r}(x+y)]p'(x+y) \nonumber
\\ & \qquad  -x^\alpha \bar{r}(x+y)p''(x+y) -k[2 \alpha x^{\alpha-1} p'(x+y)+x^\alpha p''(x+y)] \nonumber
\\ & \qquad + \alpha(\alpha-1)x^{\alpha-2}[\bar{r}(x+y)(1-p(x+y)) - kp(x+y)] < 0,
\label{eq:prospect_secondderivative}
\end{align}
i.e., $\mathbb{E}(u)$ is strictly concave. Note that $0 < \alpha \leq 1$ and $f(x+y)>0$. We now argue there is a single interval where $[x\bar{r}'(x+y)+\alpha \bar{r}(x+y)]>0$. 

As $\bar{r}$ is concave and decreasing, $[x \bar{r}'(x+y)+\alpha \bar{r}(x+y)]$ is decreasing in $x$. For $y<\bar{y}$ and $x$ sufficiently close to $0$, we have $[x\bar{r}'(x+y)+\alpha \bar{r}(x+y)] > 0$. Let us define $c:= \sup\{x \in [0,1] | x\bar{r}'(x+y)+\alpha \bar{r}(x+y) > 0\}$. Thus, the only region in which $[x\bar{r}'(x+y)+\alpha \bar{r}(x+y)]$ is nonnegative is $[0,c]$. As a result, we can have exactly one maximum of $\mathbb{E}(u)$ as a root of $\frac{d\mathbb{E}(u)}{dx}$ by strict concavity, and this root must be strictly less than $c$. Hence $B(y)$ is single-valued for $y\in[0,\bar{y})$. Furthermore, when $y \geq \bar{y}$, $B(y)$ is again single-valued with value zero by Lemma~\ref{lemma:interval}.
\smallskip

\textbf{Case 2:}  {\textit{$\bar{r}(x_T)$ is increasing}.}

Recall from Lemma~\ref{lemma:interval} that $\forall x \in \mathbb{R}_{\geq 0}$ such that $x+y \in \mathcal{I}$, we have $f(x+y)>0$ and $f'(x+y)<0$. Concavity of $f$ is immediate from~\eqref{eq:effectiverorconcavity}. In this interval $\mathcal{I}$, the expected utility is concave, i.e., 
\begin{equation*}
\frac{\partial^2 \mathbb{E}(u)}{\partial x^2} = x^\alpha f''(x+y)+2\alpha x^{\alpha-1} f'(x+y) + \alpha(\alpha-1)x^{\alpha-2}f(x+y)< 0.
\end{equation*}
Since any best response $b(y) \in B(y)$ satisfies $b(y)+y \in \mathcal{I}$, $b(y)$ must be a maximum of $\mathbb{E}(u)$, and therefore it is unique. 
\qed

\vspace{4mm}
The proof of Lemma~\ref{lemma:CDcontinuity} is based on Berge's Maximum theorem, which we state below. 

\begin{theorem}[\cite{ok2007real}]
Let $\Theta$ and $X$ be two metric spaces, and $C:\Theta \rightrightarrows X$ a compact-valued correspondence. Let the function $\Phi: X \times \Theta \to \mathbb{R}$ be jointly continuous in both $X$ and $\Theta$. Define
\begin{align*}
\sigma(\theta) &:= \argmax_{x \in C(\theta)} \Phi(x,\theta), 
\text{ and }
\\ \Phi^*(\theta) &:= \max_{x \in C(\theta)} \Phi(x,\theta), \forall \theta \in \Theta.
\end{align*}
If $C$ is continuous at $\theta \in \Theta$, then
\begin{enumerate}
\item $\sigma:\Theta \rightrightarrows X$ is compact-valued, upper hemicontinuous and closed at $\theta$. 
\item $\Phi^*:\Theta \rightarrow \mathbb{R}$ is continuous at $\theta$. 
\end{enumerate}
\label{theorem:berge}
\end{theorem}

\noindent \textbf{Proof of Lemma~\ref{lemma:CDcontinuity}.}
The best response correspondence of player $i$, $B_i:\bar{S}_{-i} \rightrightarrows S_i$ is analogous to $\sigma$ according to the notation in Theorem~\ref{theorem:berge}, and the expected utility $\mathbb{E}(u)$ is analogous to the function $\Phi$. Since we compute the best response of a player in its entire domain $[0,1]$, we define the correspondence $C:\bar{S}_{-i} \rightrightarrows [0,1]$ for any joint strategies of players other than $i$. Therefore, $C$ is compact-valued, and both upper and lower hemicontinuous. Hence $B_i$ is upper hemicontinuous (from Theorem~\ref{theorem:berge}), and as it is single-valued (from Lemma~\ref{lemma:CDuniquebr}), $B_i$ is continuous. The result holds for every player $i \in \mathcal{N}$ by symmetry.
\qed

\vspace{4mm}

\noindent \textbf{Proof of Lemma~\ref{lemma:garg}.}
Recall that $g(x_T)=-\frac{\alpha f(x_T)}{f'(x_T)}$. Then,
\begin{align}
\frac{1}{\alpha}\frac{\partial g(x_T)}{\partial x}  &=  -\frac{(f'(x_T))^2 - f(x_T)f''(x_T)}{(f'(x_T))^2}. \label{eq:appendix}
\end{align}

\textbf{Case 1:} {\textit{$r(x_T)$ is decreasing}.}

For clarity of presentation, denote $f(x_T),\bar{r}(x_T)$ and $p(x_T)$ as $f,\bar{r},p$, respectively. We have
\begin{align*}
f & =\bar{r}(1-p)-kp,
\\ f'& =\bar{r}'(1-p)-\bar{r}p'-kp',
\\ f''& =\bar{r}''(1-p)-2\bar{r}'p'-\bar{r}p''-kp'',
\\ f'^2 & =(\bar{r}'(1-p))^2+(p'(\bar{r}+k))^2 -2\bar{r}'p'(\bar{r}+k)(1-p),
\\ ff'' & =(\bar{r}(1-p)-kp)(\bar{r}''(1-p)-\bar{r}p''-kp'') +  2\bar{r}'p'kp-2\bar{r}'p'\bar{r}(1-p).
\end{align*}

Note that $f>0$ for $x_T \in \mathcal{I}$. Let $\rho = \bar{r}''(1-p)-\bar{r}p''-kp'' \leq 0$. Then,
\begin{align*}
ff'' & =f\rho+2\bar{r}'p'kp-2\bar{r}'p'\bar{r}(1-p),
\\ f'^2-ff'' &= (\bar{r}'(1-p))^2+(p'(\bar{r}+k))^2 - 2\bar{r}'p'(\bar{r}+k)(1-p) - f\rho -2\bar{r}'p'kp+2\bar{r}'p'\bar{r}(1-p)
\\ &= (\bar{r}'(1-p))^2+(p'(\bar{r}+k))^2 - f\rho -2\bar{r}'p'k > 0.
\end{align*}

Therefore, $\frac{\partial g(x_T)}{\partial x_T} <0$. 

\vspace{2mm}

\textbf{Case 2:} {\textit{$r(x_T)$ is increasing}.}

When $x_T \in \mathcal{I}$, $f(x_T)>0$. Recall from the proof of Lemma~\ref{lemma:interval} that $f''(x_T) \leq 0$ for all $x_T \in \mathcal{I}$. Thus, it follows directly from~\eqref{eq:appendix} that $\frac{\partial g(x_T)}{\partial x_T} <0$.  
\qed

\vspace{3mm}

\noindent \textbf{Proof of Theorem~\ref{theorem:pnecharacterization}.}

\noindent \textit{Existence of PNE:} We first prove the existence of (at least one) PNE using Brouwer's fixed point theorem. 

The strategy set of each player $i$, $S_i$, is a convex and compact subset of the Euclidean space, and so is the joint strategy space, $S \subset \mathbb{R^{|\mathcal{N}|}}$. Define a mapping $T: S \to S$ such that $T(x_1,\ldots,x_{|\mathcal{N}|}) = (B_1(y_1),\ldots,B_{|\mathcal{N}|}(y_{|\mathcal{N}|}))$, where $y_i = \sum_{j=1,j \neq i}^{|\mathcal{N}|} x_j$. From Lemma~\ref{lemma:CDuniquebr}, $T$ is well defined and unique over the entire domain, and from Lemma~\ref{lemma:CDcontinuity}, it is continuous. Brouwer's fixed point theorem then guarantees the existence of a strategy profile $s=\{x^*_i\}_{i \in \mathcal{N}} \in S$ that is invariant under the best response mapping, and therefore is a PNE of the game.

\vspace{2mm}

\noindent \textit{Uniqueness of PNE:} We next prove the uniqueness of PNE using Lemma~\ref{lemma:garg}. 

Suppose the claim is false, and consider two PNEs of the game with total investments $x^*_{T1}$ and $x^*_{T2}$, respectively. Let $x^*_{T2} > x^*_{T1}$, without loss of generality. Let the respective supports of the PNEs be denoted as $\mathcal{S}_1$ and $\mathcal{S}_2$. Clearly from the definition of the support in \eqref{eq:supportdef}, we have $\mathcal{S}_2 \subseteq \mathcal{S}_1$. We know from~\eqref{eq:focbestresponse} that the investment of a player $i$ in the support of a PNE with total investment $x^*_{T}$ is $g_i(x^*_{T})$. Thus for both PNEs considered here, we have
\begin{align*}
& \sum_{j \in \mathcal{S}_1} g_j(x^*_{T1}) = x^*_{T1}, \quad \sum_{j \in \mathcal{S}_2} g_j(x^*_{T2}) = x^*_{T2}.
\end{align*} 
We can rewrite the first identity above as
\begin{align*}
& \sum_{j \in \mathcal{S}_2} g_j(x^*_{T1}) + \sum_{j \in \mathcal{S}_1\setminus \mathcal{S}_2} g_j(x^*_{T1}) = x^*_{T1}
\\ \implies & \sum_{j \in \mathcal{S}_2} g_j(x^*_{T1}) \leq x^*_{T1} < x^*_{T2} = \sum_{j \in \mathcal{S}_2} g_j(x^*_{T2}). 
\end{align*} 
However, from Lemma~\ref{lemma:garg} we know that if $x^*_{T2} > x^*_{T1}$, we have
$$  g_j(x^*_{T1})  > g_j(x^*_{T2}), \forall j \in \mathcal{S}_2, $$
which is a contradiction. Thus $x^*_{T1} = x^*_{T2}$, and both PNEs have identical support. 

Now, for a given total investment $x^*_{T}$ at a PNE, the optimal investment by the players is unique, which follows again from strict monotonicity of $g_j(x^*_{T})$ for every player in the support. This concludes the proof. 
\qed

\section{Convergence to Nash equilibrium}
\label{appendix:wsts}

We first observe that Fragile CPR games belong to a class of games known as \textit{Weak Strategic Substitutes (WSTS) games with aggregation}~\cite{dubey2006strategic}. The definition is reproduced here. 

\begin{definition}\label{def:wsts}
A strategic game $\Gamma(\mathcal{N},\{S_i\}_{i \in \mathcal{N}},\{u_i\}_{i \in \mathcal{N}})$ is a \textit{Weak Strategic Substitutes} game with aggregation if the player utilities are defined as $u_i: S_i \times \bar{S}_{-i} \to \mathbb{R}$ and there exists a function $b_i:\bar{S}_{-i}\to \mathbb{R}$ for every $i \in \mathcal{N} $such that:
\begin{enumerate}
\item $b_i(y) \in \argmaxl_{x \in S_i} u_i(x,y)$,  $\forall y \in \bar{S}_{-i}$, 
\item $b_i$ is continuous on $\bar{S}_{-i}$ and 
\item for $y_1,y_2 \in \bar{S}_{-i}$ and $y_1>y_2$, $b_i(y_1) \leq b_i(y_2)$.
\end{enumerate}
\end{definition}

Note that Lemma~\ref{lemma:CDuniquebr} and Lemma~\ref{lemma:CDcontinuity} verify the first two properties in Definition~\ref{def:wsts}, respectively. The monotonicity of the best response also holds as a consequence of Lemma~\ref{lemma:garg} as we show below. 

\begin{proposition} In a Fragile CPR game, the best response $B(y)$ is decreasing in $y \in \bar{S}$.
\label{proposition:CDmonotonicbr}
\end{proposition}

\begin{proof}
Let $x_1=B(y_1)$ and $x_2=B(y_2)$, where $y_1,y_2 \in \bar{S}$ with $y_2>y_1$. Let $y_2 < \bar{y}$ (recall that $f(\bar{y})=0$), as the result holds trivially otherwise. According to the definition of $g(x_T)$ given in~\eqref{eq:focbestresponse}, we have $x_1 = g(x_1+y_1)$ and $x_2=g(x_2+y_2)$. Recall from the proof of Lemma~\ref{lemma:interval} that $B(y)+y \in \mathcal{I}$. Thus, if $x_2>x_1$, it follows from Lemma~\ref{lemma:garg} that $x_2 = g(x_2+y_2) < g(x_1+y_1) = x_1$, which is a contradiction. 
\end{proof}

Thus, Fragile CPR games are instances of WSTS games. Furthermore, since the best response correspondence of each player is single-valued (by Lemma~\ref{lemma:CDuniquebr}), it follows from Remark 2 in~\cite{dubey2006strategic} that Fragile CPR games are instances of \textit{best-response potential games}. Thus, for each Fragile CPR game, there exists a potential function, continuous in players' strategies and the functions $b_i$'s, whose maximum corresponds to the PNE of the game. As a consequence, sequential best response dynamics~\cite{jensen2006stability} and simultaneous better reply dynamics~\cite{dubey2006strategic} converge to the PNE. Note that a player need not know the risk preferences of other players to compute her best response. Knowledge of her own risk preferences and the aggregate investment by other players is sufficient for convergence to PNE under the sequential best response dynamics. 

Note that the existence of a continuous best response potential function directly implies the existence of a PNE from the extreme value theorem. Therefore, this gives an alternative proof of existence of PNE in Fragile CPR games. However to establish that the function defined in~\cite{dubey2006strategic} is a best-response potential function, one needs to first verify certain continuity and monotonicity properties of the best response correspondence. Thus our first principles analysis yields a simpler and more direct proof of the result. We further show the uniqueness of the PNE from the structural properties of the best response (Lemma~\ref{lemma:garg}), which is not established immediately from using the potential function. 

\section{Proofs pertaining to fragility under competition}
\label{appendix:fucproof}

\noindent \textbf{Proof of Proposition~\ref{proposition:boundedpneXT}.}
First we show that $\{x^*_{Tn}\}$ is a monotonically increasing sequence. 

Since $x^*_{Tn}$ satisfies~\eqref{eq:homogeneous_foc_PNE}, we have $x^*_{Tn} \in \mathcal{I}, \forall n \in \mathbb{N}$ (the interval $\mathcal{I}$ as defined in Lemma~\ref{lemma:interval}). Let $n,m \in \mathbb{N}$ be such that $n < m$, and assume on the contrary that $x^*_{Tn} \geq x^*_{Tm}$. However, from~\eqref{eq:homogeneous_foc_PNE} and Lemma~\ref{lemma:garg} we have
\begin{align*}
& \frac{x^*_{Tn}}{n} = \frac{\alpha f(x^*_{Tn})}{(-f'(x^*_{Tn}))} \leq \frac{\alpha f(x^*_{Tm})}{(-f'(x^*_{Tm}))} = \frac{x^*_{Tm}}{m}.
\end{align*}
This implies that $x^*_{Tn} < x^*_{Tm}$, yielding the desired contradiction. Thus the total investment at the PNE is strictly increasing in the number of players. Since $p(x_T)$ is strictly increasing in $x_T$, the fragility is strictly increasing in the number of players as well. 

Next we show that the sequence $\{x^*_{Tn}\}$ converges to $\bar{y}$. Recall from the proof of Lemma~\ref{lemma:interval} that $\bar{y} \in [0,1)$ is defined such that $f(\bar{y})=0$ and $f'(\bar{y})<0$. Since we only consider games with nonzero investment at the PNE, we have $x^*_{Tn} < \bar{y}, \forall n\in\mathbb{N}$; otherwise the effective rate of return will be nonpositive, and players will prefer to invest entirely in the safe resource. Thus $\{x^*_{Tn}\}$ converges since it is a monotonically increasing and bounded sequence of real numbers.

Let $y=\bar{y}-\epsilon$ for some $\epsilon>0$ such that $y \in \mathcal{I}$. Since $f(y)>0$, we can choose $n_0$ large enough such that
$$ \frac{y}{n_0} f'(y) + \alpha f(y) > 0.$$
We will show that $y < x^*_{Tn_0}$. Assume to the contrary that $y \geq x^*_{Tn_0}$. Since both $y, x^*_{Tn_0} \in \mathcal{I}$, we have
\[ \frac{x^*_{Tn_0}}{n_0} = \frac{\alpha f(x^*_{Tn_0})}{(-f'(x^*_{Tn_0}))} \geq \frac{\alpha f(y)}{(-f'(y))} > \frac{y}{n_0}. \]
The first equality is due to~\eqref{eq:homogeneous_foc_PNE}, and the first inequality is a consequence of Lemma~\ref{lemma:garg}. Thus, we have the required contradiction. For any arbitrarily small $\epsilon>0$, we can always find $n_0 \in \mathbb{N}$ large enough such that $\bar{y}-\epsilon < x^*_{Tn} < \bar{y}, \forall n \geq n_0$, and thus $\{x^*_{Tn}\}$ converges to $\bar{y}$. 
\qed

\vspace{3mm}

\noindent \textbf{Proof of Lemma~\ref{lemma:gammahat}.} Let $x_\mathtt{PVT}$ be the optimal investment by a single player in the CPR of $\Gamma$. We define a tangent to $\bar{r}(x_T)=(r(x_T)-1)^\alpha$ at $x_\mathtt{PVT}$, i.e., let
\begin{equation}
\hat{r}(x_T) \triangleq \bar{r}(x_\mathtt{PVT}) + (x_T-x_\mathtt{PVT}) \bar{r}'(x_\mathtt{PVT}).
\label{eq:inefficiency_tilderor}
\end{equation}
Denote by $\hat{\Gamma}$ the perturbed Fragile CPR game with rate of return $1+\hat{r}^{\frac{1}{\alpha}}$, and all other parameters the same as in $\Gamma$. Note that $\hat{r}(x_T)$ is affine and monotonic, and $\hat{r}(x_T) \geq \bar{r}(x_T) > 0, \forall x_T \in [0,1]$, and thus satisfies Assumption~\ref{assumption:CDCI}. 

Let $\hat{x}_\mathtt{PVT}$ be the optimal investment by a single player in the CPR of the perturbed game $\hat{\Gamma}$. First we show that $\hat{x}_\mathtt{PVT} = x_\mathtt{PVT}$. Let $\hat{f}(x_T) := \hat{r}(x_T) (1-p(x_T)) - kp(x_T)$ be the effective rate of return of the player in $\hat{\Gamma}$. From the definition of $\hat{r}$, we have $\hat{f}(x_\mathtt{PVT}) = f(x_\mathtt{PVT})$ and $\hat{f'}(x_\mathtt{PVT}) = f'(x_\mathtt{PVT})$. Therefore $x_\mathtt{PVT}$ satisfies the necessary condition of optimal investment by a single player in $\hat{\Gamma}$ given by
\begin{equation}\label{eq:focgammahat1}
x_\mathtt{PVT}\hat{f}'(x_\mathtt{PVT})+\alpha \hat{f}(x_\mathtt{PVT})=0.
\end{equation}
From the uniqueness of optimal investment (Lemma~\ref{lemma:CDuniquebr}), we must have $\hat{x}_\mathtt{PVT} = x_\mathtt{PVT}$.

Now, it suffices to show that $\hat{\bar{y}} \geq \bar{y}$, where $\bar{y}, \hat{\bar{y}}$ are as defined in Lemma~\ref{lemma:interval} for effective rate of return functions $f$ and $\hat{f}$, respectively. From the concavity of $\bar{r}(x_T)$, it follows that $\hat{r}(x_T) \geq \bar{r}(x_T)$. Since $f(\bar{y})=0$, we must have $\hat{f}(\bar{y}) \geq 0$. From the definition of $\hat{\bar{y}}$, we know that $\hat{f}(y) < 0$ when $y > \hat{\bar{y}}$. As a result, we have $\hat{\bar{y}} \geq \bar{y}$. 
\qed

\vspace{3mm}

\noindent \textbf{Proof of Theorem~\ref{theorem:FuCdecreasing}.}
We know from Lemma~\ref{lemma:gammahat} that it suffices to consider rate of return functions of the form $r(x_T) = 1+(ax_T+b)^{\frac{1}{\alpha}}, a \leq 0, b > 0$ to establish the result. Thus, we consider effective rate of return functions of the form $f(x_T) = (ax_T+b)(1-p(x_T))-kp(x_T)$.  

Define $\lambda \triangleq (1+\frac{2}{\alpha})$. Since $f(x_T) > 0$ for $x_T \in [x_\mathtt{PVT},\bar{y})$ and $f(x_T)<0$ for $x_T > \bar{y}$, it suffices to show that $f(\lambda x_\mathtt{PVT}) < 0$. Suppose $\lambda x_\mathtt{PVT} < 1$, for otherwise the result holds trivially. First we compute,
\begin{align}
f(\lambda x_\mathtt{PVT}) - f(x_\mathtt{PVT}) &= (a\lambda x_\mathtt{PVT} + b) (1-p(\lambda x_\mathtt{PVT})) - k p(\lambda x_\mathtt{PVT}) - (ax_\mathtt{PVT}+b)(1-p(x_\mathtt{PVT})) + kp(x_\mathtt{PVT}) \nonumber
\\ & = ax_\mathtt{PVT} (\lambda-1) (1-p(x_\mathtt{PVT})) + (ax_\mathtt{PVT}\lambda+b+k)(p(x_\mathtt{PVT})-p(\lambda x_\mathtt{PVT})) \nonumber
\\ & = ax_\mathtt{PVT} \frac{2}{\alpha} (1-p(x_\mathtt{PVT})) + (ax_\mathtt{PVT}\lambda+b+k)(p(x_\mathtt{PVT})-p(\lambda x_\mathtt{PVT})). \label{eq:differencef}
\end{align} 
In the last step, we have used $\lambda-1=\frac{2}{\alpha}$.

Recall that $x_\mathtt{PVT}$ satisfies the first order condition, $x_\mathtt{PVT}f'(x_\mathtt{PVT})+\alpha f(x_\mathtt{PVT}) = 0$. Using~\eqref{eq:differencef}, we have
\begin{align*}
\alpha f(\lambda x_\mathtt{PVT}) & = \alpha [f(\lambda x_\mathtt{PVT}) - f(x_\mathtt{PVT})] - x_\mathtt{PVT}f'(x_\mathtt{PVT}) 
\\ & = 2ax_\mathtt{PVT} (1-p(x_\mathtt{PVT})) + \alpha(ax_\mathtt{PVT}\lambda+b+k)(p(x_\mathtt{PVT})-p(\lambda x_\mathtt{PVT})) 
\\ & \quad - x_\mathtt{PVT} [a (1-p(x_\mathtt{PVT})) - (ax_\mathtt{PVT}+b+k)p'(x_\mathtt{PVT})]
\\ & = ax_\mathtt{PVT} (1-p(x_\mathtt{PVT})) + \alpha(ax_\mathtt{PVT}\lambda+b+k)(p(x_\mathtt{PVT})-p(\lambda x_\mathtt{PVT})) + \alpha (ax_\mathtt{PVT}+b+k) \frac{x_\mathtt{PVT}}{\alpha} p'(x_\mathtt{PVT})
\\ & = ax_\mathtt{PVT} (1-p(x_\mathtt{PVT})) + \alpha(ax_\mathtt{PVT}\lambda+b+k)(p(x_\mathtt{PVT})-p(\lambda x_\mathtt{PVT})) 
\\ & \quad + \alpha (a\lambda x_\mathtt{PVT}+b+k) \frac{x_\mathtt{PVT}}{\alpha} p'(x_\mathtt{PVT}) + (1-\lambda) a(x_\mathtt{PVT})^2 p'(x_\mathtt{PVT}) 
\\ & = ax_\mathtt{PVT} \left[1-p(x_\mathtt{PVT}) - \frac{2}{\alpha} x_\mathtt{PVT} p'(x_\mathtt{PVT})\right] + \alpha (a\lambda x_\mathtt{PVT}+b+k) \left[p(x_\mathtt{PVT})-p(\lambda x_\mathtt{PVT}) + \frac{x_\mathtt{PVT}}{\alpha} p'(x_\mathtt{PVT})\right] 
\\ & \leq ax_\mathtt{PVT}(1-p(\lambda x_\mathtt{PVT})) + \alpha (a\lambda x_\mathtt{PVT}+b+k) \left[ p\left(x_\mathtt{PVT}\left(1+\frac{1}{\alpha}\right)\right)-p(\lambda x_\mathtt{PVT})\right] < 0. 
\end{align*}
In the second last step we have used the following inequalities which follow from the convexity of $p$:
\begin{align*}
p(x_\mathtt{PVT}) + \frac{2}{\alpha} x_\mathtt{PVT} p'(x_\mathtt{PVT}) & \leq p\left(x_\mathtt{PVT}\left(1+\frac{2}{\alpha}\right)\right), \text{  and  } p(x_\mathtt{PVT}) + \frac{1}{\alpha} x_\mathtt{PVT} p'(x_\mathtt{PVT}) \leq p\left(x_\mathtt{PVT}\left(1+\frac{1}{\alpha}\right)\right).
\end{align*}
The last inequalities follow as $a \leq 0$, $\lambda x_\mathtt{PVT} < 1$, $a+b>0$ and $p$ is strictly increasing. As a result we have $f(\lambda x_\mathtt{PVT}) < 0$, and therefore 
$\bar{y} < x_\mathtt{PVT}\left(1+\frac{2}{\alpha}\right)$. 
\qed

\vspace{3mm}

\noindent \textbf{Proof of Corollary~\ref{corollary:FuCdecreasing}.}
Suppose $p(x) = e_0+e_1x+e_2x^2+\ldots+e_\gamma x^\gamma$, with $e_i \geq 0, i \in \{0,1,2,\ldots,\gamma-1\}$ and $e_\gamma > 0$. Then for $\lambda=1+\frac{2}{\alpha}>1$, we have
$$ p(\lambda x) = \sum_{i=0}^\gamma e_i \lambda^i x^i \leq \lambda^\gamma \sum_{i=0}^\gamma e_ix^i = \lambda^\gamma p(x). $$
From Theorem~\ref{theorem:FuCdecreasing} we know that for decreasing rate of return functions, $\bar{y} < \lambda x_\mathtt{PVT}$. Therefore, 
$$ \mathcal{F}(\Gamma) \leq \frac{p(\bar{y})}{p(x_\mathtt{PVT})} < \left(1+\frac{2}{\alpha}\right)^\gamma. $$
This concludes the proof. 
\qed

\vspace{3mm}

\noindent \textbf{Proof of Theorem~\ref{theorem:fuclowerbound}.} We first analyze the behavior of $p(\bar{y})$. Note that $\bar{y}$ is defined as the solution to $f(x) = 0$, i.e.,
\begin{align}
& \bar{r}(\bar{y})(1-p(\bar{y})) - kp(\bar{y}) = 0 \nonumber
\\ \implies & p(\bar{y}) = \frac{\bar{r}(\bar{y})}{\bar{r}(\bar{y})+k} > \frac{\bar{r}(1)}{\bar{r}(1)+k}, \label{eq:decreasing_ybar1}
\end{align}
where the inequality holds because $k>0$ and $\bar{r}(x_T)$ is strictly decreasing in $x_T$ with $\bar{r}(1)>0$ from our assumptions. As a result, $p(\bar{y}) = \bar{y}^\gamma$ is bounded away from $0$. Next we obtain an upper bound on $x_\mathtt{PVT}$. We compute $x_Tf'(x_T)+\alpha f(x_T)$ as 
\begin{align}
x_Tf'(x_T)+\alpha f(x_T) & = x_T[\bar{r}'(x_T)(1-p(x_T)) - (\bar{r}(x_T)+k)p'(x_T)] + \alpha[\bar{r}(x_T)(1-p(x_T))-kp(x_T)] \nonumber
\\ & = [x_T\bar{r}'(x_T)+\alpha \bar{r}(x_T)](1-p(x_T)) - (\bar{r}(x_T)+k) x_T p'(x_T) -\alpha kp(x_T) \label{eq:privatefoc12}
\\ & = x_T\bar{r}'(x_T)+\alpha \bar{r}(x_T) - x_T\bar{r}'(x_T)p(x_T) - (\bar{r}(x_T)+k)(x_Tp'(x_T)+\alpha p(x_T)). \label{eq:privatefoc11}
\end{align}
By the definition of $x_\mathtt{PVT}$, \eqref{eq:privatefoc11} evaluates to $0$ at $x_T=x_\mathtt{PVT}$. 

First we consider the case where $x^*_r < 1$. From our assumptions, $x_T\bar{r}'(x_T)+\alpha \bar{r}(x_T)$ is strictly decreasing in $x_T$. Since $x^*_r$ is the maximizer of $x^\alpha_T\bar{r}(x_T)$ for $x_T \in [0,\infty)$, we have $x^*_r\bar{r}'(x^*_r)+\alpha\bar{r}(x^*_r) = 0$. Furthermore, at any $x_T > x^*_r$, $x_T\bar{r}'(x_T)+\alpha \bar{r}(x_T) < 0$. 

Note that in \eqref{eq:privatefoc12}, the second and third terms are nonpositive for every $x_T \in [0,1]$ following our assumptions. Therefore, at every $x_T \in (x^*_r,1]$, $x_Tf'(x_T)+\alpha f(x_T)$ is strictly negative. Thus, we must have $x_\mathtt{PVT} \leq x^*_r$, which yields the following lower bound:

$$ \frac{p(\bar{y})}{p(x_\mathtt{PVT})} > \frac{\bar{r}(1)}{\bar{r}(1)+k} \frac{1}{(x^*_r)^\gamma}. $$

Next suppose that $x^*_r > 1$. From \eqref{eq:decreasing_ybar1} we obtain 
\begin{equation}
p(\bar{y}) = \frac{\bar{r}(\bar{y})}{\bar{r}(\bar{y})+k} < \frac{\bar{r}(0)}{\bar{r}(0)+k}. \label{eq:decreasing_ybar_temp}
\end{equation}
From the first order condition of optimality, \eqref{eq:privatefoc11} evaluates to $0$ at $x_T=x_\mathtt{PVT}$, i.e.,
\begin{align}
[x_\mathtt{PVT}\bar{r}'(x_\mathtt{PVT})+\alpha \bar{r}(x_\mathtt{PVT})] & - x_\mathtt{PVT}\bar{r}'(x_\mathtt{PVT})p(x_\mathtt{PVT}) - (\bar{r}(x_\mathtt{PVT})+k)(x_\mathtt{PVT}p'(x_\mathtt{PVT})+\alpha p(x_\mathtt{PVT})) = 0 \label{eq:decreasing_pvt_foc1}
\\ \implies x_\mathtt{PVT}\bar{r}'(x_\mathtt{PVT})+\alpha \bar{r}(x_\mathtt{PVT}) & = x_\mathtt{PVT}\bar{r}'(x_\mathtt{PVT})p(x_\mathtt{PVT}) + (\bar{r}(x_\mathtt{PVT})+k)(x_\mathtt{PVT}p'(x_\mathtt{PVT})+\alpha p(x_\mathtt{PVT})) \nonumber
\\ & \leq (x_\mathtt{PVT}\bar{r}'(x_\mathtt{PVT}) + (\bar{r}(x_\mathtt{PVT})+k) (\alpha+\zeta)) p(x_\mathtt{PVT}) \qquad \text{(from the definition of $\zeta$)} \nonumber
\\ \implies p(x_\mathtt{PVT}) & \geq \frac{x_\mathtt{PVT}\bar{r}'(x_\mathtt{PVT})+\alpha \bar{r}(x_\mathtt{PVT})}{x_\mathtt{PVT}\bar{r}'(x_\mathtt{PVT}) + (\bar{r}(x_\mathtt{PVT})+k)(\alpha+\zeta)} \nonumber
\\ & = \left[1+\frac{\zeta\bar{r}(x_\mathtt{PVT})+k(\alpha+\zeta)}{x_\mathtt{PVT}\bar{r}'(x_\mathtt{PVT})+\alpha \bar{r}(x_\mathtt{PVT})}\right]^{-1} \nonumber
\\ & \geq \left[1+\frac{\zeta\bar{r}(0)+k(\alpha+\zeta)}{\bar{r}'(1)+\alpha \bar{r}(1)}\right]^{-1}. \label{eq:decreasing_xpvt_temp}
\end{align}
The last two inequalities hold as a consequence of $x^*_r > 1$, which implies $x_T\bar{r}'(x_T)+\alpha \bar{r}(x_T) > 0$ for $x_T \in [x_\mathtt{PVT},1]$ and is monotonically decreasing. From \eqref{eq:decreasing_ybar_temp} and \eqref{eq:decreasing_xpvt_temp}, we conclude, 
\begin{equation*}
\mathcal{F}(\Gamma) \leq \frac{p(\bar{y})}{p(x_\mathtt{PVT})} < \frac{\bar{r}(0)}{\bar{r}(0)+k} \left[1+\frac{\zeta(\bar{r}(0)+k)+k\alpha}{\bar{r}'(1)+\alpha \bar{r}(1)}\right].
\end{equation*}
This concludes the proof. 
\qed

\vspace{3mm}

\noindent \textbf{Proof of Proposition~\ref{proposition:decreasing}.} 
The bound obtained in Theorem~\ref{theorem:FuCdecreasing} is a stronger bound when 
\begin{align*}
& \left(1+\frac{2}{\alpha}\right) \leq \frac{1}{p(x_\mathtt{PVT})} \implies p(x_\mathtt{PVT}) = x_\mathtt{PVT} \leq \left(\frac{\alpha}{\alpha+2}\right).
\end{align*}
A necessary and sufficient condition for the above equation to be true is that $x_Tf'(x_T)+\alpha f(x_T) \leq 0$ at $x_T = \left(\frac{\alpha}{\alpha+2}\right) = p(x_T)$. Following \eqref{eq:privatefoc11}, we evaluate $x_Tf'(x_T)+\alpha f(x_T)$ at $x_T = \left(\frac{\alpha}{\alpha+2}\right)$ and obtain
\begin{align*}
 x_Tf'(x_T)+\alpha f(x_T) & =  x_T\bar{r}'(x_T)(1-p(x_T)) + \alpha \bar{r}(x_T) - (\bar{r}(x_T)+k)(x_Tp'(x_T)+\alpha p(x_T))
\\ & =  x_T\bar{r}'(x_T)\left[1-\frac{\alpha}{\alpha+2}\right] + \alpha \bar{r}(x_T) - (\bar{r}(x_T)+k) (\alpha+1) p(x_T)
\\ & =   x_T\bar{r}'(x_T) \left[\frac{2}{\alpha+2}\right] + \bar{r}(x_T) \left[\alpha - (\alpha+1) \frac{\alpha}{\alpha+2}\right] - k \frac{\alpha(\alpha+1)}{\alpha+2}
\\ & =  x_T\bar{r}'(x_T) \left[\frac{2}{\alpha+2}\right] + \bar{r}(x_T) \left[\frac{\alpha}{\alpha+2}\right] - k \frac{\alpha(\alpha+1)}{\alpha+2}.
\end{align*}
Thus, the bound $\left(1+\frac{2}{\alpha}\right)$ is tighter than the trivial bound if and only if 
$$ 2x_T\bar{r}'(x_T)+\alpha \bar{r}(x_T) \leq k\alpha(1+\alpha), \qquad \text{at} \qquad x = \frac{\alpha}{\alpha+2}. $$
When $\bar{r}(x_T)=ax_T+b$, the above condition is equivalent to
\begin{align*}
& 2\frac{\alpha}{\alpha+2}a+\alpha \left(a \frac{\alpha}{\alpha+2}+ b\right) \leq k\alpha(1+\alpha) \iff a+b \leq k(1+\alpha). 
\end{align*} 
This concludes the proof. 
\qed

\vspace{3mm}

\noindent \textbf{Proof of Theorem~\ref{theorem:FuCincreasing}.}
Following Lemma~\ref{lemma:gammahat}, we consider rate of return functions of the form $r(x_T) = 1+(ax_T+b)^{\frac{1}{\alpha}}$ with $a \geq 0$ and $b > 0$. We prove both parts of the theorem in the order they are presented. 
 
\vspace{1mm}

\textbf{Part 1:} \textit{$\frac{\bar{y}}{x_\mathtt{PVT}} \leq 1+\frac{1}{\alpha}$.} 

Let $x_\mathtt{PVT}$ be the optimal investment by a single player in the resource. Assume that $x_\mathtt{PVT}(1+\frac{1}{\alpha}) < 1$, otherwise the result follows since $\bar{y} \leq 1$. In addition, we have $x_\mathtt{PVT}(1+\frac{1}{1+\alpha}) < 1$. Define $\mu \triangleq (1+\frac{1}{\alpha})$ for ease of exposition. We will use the following two inequalities which are due to the convexity and monotonicity of the failure probability function $p$:
\begin{align}
x_\mathtt{PVT} p'(x_\mathtt{PVT}) + (1+\alpha) p(x_\mathtt{PVT}) & = (1+\alpha) \left[ \frac{x_\mathtt{PVT}}{1+\alpha} p'(x_\mathtt{PVT}) + p(x_\mathtt{PVT}) \right] \leq (1+\alpha) p\left(x_\mathtt{PVT} \left(1+\frac{1}{1+\alpha}\right)\right) \nonumber
\\ & \leq (1+\alpha) p(\mu x_\mathtt{PVT}) \label{eq:convexpplusone},
\\ x_\mathtt{PVT} p'(x_\mathtt{PVT}) + \alpha p(x_\mathtt{PVT}) & \leq \alpha p(\mu x_\mathtt{PVT}) \label{eq:convexp}.
\end{align}

At the optimal investment by a single player, we have
\begin{align*}
0 & = x_\mathtt{PVT}f'(x_\mathtt{PVT})+\alpha f(x_\mathtt{PVT}) 
\\ & = x_\mathtt{PVT}\left[a(1-p(x_\mathtt{PVT})) -ax_\mathtt{PVT}p'(x_\mathtt{PVT})-(b+k)p'(x_\mathtt{PVT})\right] + \alpha \left[ax_\mathtt{PVT}-ax_\mathtt{PVT}p(x_\mathtt{PVT})+b-(b+k)p(x_\mathtt{PVT})\right]
\\ & = -ax_\mathtt{PVT}\left[x_\mathtt{PVT}p'(x_\mathtt{PVT})+(1+\alpha)p(x_\mathtt{PVT})\right] - (b+k)\left[x_\mathtt{PVT}p'(x_\mathtt{PVT})+\alpha p(x_\mathtt{PVT})\right] + ax_\mathtt{PVT}(1+\alpha) + \alpha b
\\ & \geq -ax_\mathtt{PVT}(1+\alpha)p(\mu x_\mathtt{PVT}) - (b+k) \alpha p(\mu x_\mathtt{PVT}) +  ax_\mathtt{PVT}(1+\alpha) + \alpha b \text{\quad\quad(from~\eqref{eq:convexpplusone} and~\eqref{eq:convexp})}
\\ & = \alpha \left([a \mu x_\mathtt{PVT}+b][1-p(\mu x_\mathtt{PVT})] - kp(\mu x_\mathtt{PVT})\right) 
\\ & = \alpha f(\mu x_\mathtt{PVT}).
\end{align*}

Since $f(x_T) > 0$ for $x_T \in [x_\mathtt{PVT},\bar{y})$, we have $\bar{y}\leq x_\mathtt{PVT}\left(1+\frac{1}{\alpha}\right)$. 

\vspace{2mm}

\textbf{Part 2:} \textit{$\frac{p(\bar{y})}{p(x_\mathtt{PVT})} \leq 1+\frac{\zeta}{\alpha}$.} 

\vspace{2mm}

We again start from the first order condition of optimality at $x_\mathtt{PVT}$. We have
\begin{align}
0 & = x_\mathtt{PVT}f'(x_\mathtt{PVT})+\alpha f(x_\mathtt{PVT}) \nonumber
\\ & = x_\mathtt{PVT} [a(1-p(x_\mathtt{PVT}))-p'(x_\mathtt{PVT})(ax_\mathtt{PVT}+b+k)] + \alpha [(ax_\mathtt{PVT}+b)(1-p(x_\mathtt{PVT}))-kp(x_\mathtt{PVT})] \nonumber
\\ & = ax_\mathtt{PVT} (1-p(x_\mathtt{PVT})) + \alpha (ax_\mathtt{PVT}+b) - (\alpha p(x_\mathtt{PVT})+ x_\mathtt{PVT} p'(x_\mathtt{PVT})) (ax_\mathtt{PVT}+b+k) \nonumber
\\ & \geq ax_\mathtt{PVT} (1-p(x_\mathtt{PVT})) + \alpha (ax_\mathtt{PVT}+b) - (\alpha+\zeta)p(x_\mathtt{PVT}) (ax_\mathtt{PVT}+b+k) \text{ \qquad (by definition of $\zeta$)  } \nonumber
\\ & = ax_\mathtt{PVT} + \alpha (ax_\mathtt{PVT}+b) - p(x_\mathtt{PVT}) [ax_\mathtt{PVT} + (\alpha+\zeta)(ax_\mathtt{PVT}+b+k)] \nonumber
\\ \implies p(x_\mathtt{PVT}) & \geq \frac{ax_\mathtt{PVT} + \alpha (ax_\mathtt{PVT}+b)}{ax_\mathtt{PVT} + (\alpha+\zeta)(ax_\mathtt{PVT}+b+k)} \label{eq:pxolowerbound}.
\end{align}

Similarly at $\bar{y}$, we have,
\begin{align*}
& f(\bar{y}) = (a\bar{y}+b)(1-p(\bar{y}))-kp(\bar{y}) = 0
\\ \implies & p(\bar{y}) = \frac{a\bar{y}+b}{a\bar{y}+b+k} = 1-\frac{k}{a\bar{y}+b+k}.
\end{align*}

From Part 1 of the proof we know that $\bar{y} \leq \left(1+\frac{1}{\alpha}\right)x_\mathtt{PVT}$. Since $a \geq 0$, we have
\begin{align*}
p(\bar{y}) &= 1-\frac{k}{a\bar{y}+b+k} \leq 1-\frac{k}{a\left(1+\frac{1}{\alpha}\right)x_\mathtt{PVT}+b+k} = \frac{\alpha(ax_\mathtt{PVT}+b)+ax_\mathtt{PVT}}{\alpha(ax_\mathtt{PVT}+b+k)+ax_\mathtt{PVT}}
\\ \implies \frac{p(\bar{y})}{p(x_\mathtt{PVT})} &\leq \frac{ax_\mathtt{PVT} + (\alpha+\zeta)(ax_\mathtt{PVT}+b+k)}{\alpha(ax_\mathtt{PVT}+b+k)+ax_\mathtt{PVT}} = 1+\frac{\zeta}{\alpha+\frac{ax_\mathtt{PVT}}{ax_\mathtt{PVT}+b+k}} \leq 1+\frac{\zeta}{\alpha}.  
\end{align*}

This concludes the proof.
\qed

\vspace{3mm}

\noindent \textbf{Proof of Corollary~\ref{corollary:FuCincreasing}.}
When $p(x_T)$ is a polynomial with all nonnegative coefficients, the degree of $p(x_T)$ gives an upper bound on $\zeta$. Specifically, let $p(x_T)$ be a polynomial with degree $\gamma$ and all nonnegative coefficients, i.e., let $p(x_T) = e_0 + e_1x_T+e_2x_T^2+\ldots+e_\gamma x_T^\gamma$, with $e_i \geq 0, i \in \{0,1,2,\ldots,\gamma-1\}$ and $e_\gamma > 0$. Then, 
\begin{align*}
\frac{x_Tp'(x_T)}{p(x_T)} & = \frac{e_1x_T+2e_2x_T^2+\ldots+\gamma e_\gamma x_T^{\gamma}} {e_0 + e_1x_T+e_2x_T^2+\ldots+e_\gamma x_T^\gamma} \leq \frac{e_1x_T+2e_2x_T^2+\ldots+\gamma e_\gamma x_T^{\gamma}}{ e_1x_T+e_2x_T^2+\ldots+e_\gamma x_T^\gamma} \leq \gamma.
\end{align*}
Thus $\zeta \leq \gamma$, with equality when $p(x_T)=x_T^\gamma$. The result now directly follows from Theorem~\ref{theorem:FuCincreasing}.
\qed

\vspace{4mm}

\noindent \textbf{Proof of Proposition~\ref{proposition:increasing}.} 
We proceed in a similar manner as the proof of Proposition~\ref{proposition:decreasing}. The bound obtained in Theorem~\ref{theorem:FuCincreasing} for $p(x_T)=x^\gamma_T$ is a tighter bound when 
\begin{align*}
& \left(1+\frac{\gamma}{\alpha}\right) \leq \frac{1}{p(x_\mathtt{PVT})} \implies p(x_\mathtt{PVT}) = x_\mathtt{PVT}^\gamma \leq \left(\frac{\alpha}{\alpha+\gamma}\right).
\end{align*}
From the proof of Theorem~\ref{theorem:fuclowerbound}, we know that the first order condition of optimality in \eqref{eq:privatefoc11} must hold at $x_T=x_\mathtt{PVT}$. Since $p(0)=0$, we have $f(0) > 0$. As a result, if at a given $x_T \in [0,1]$, $x_Tf'(x_T)+\alpha f(x_T) < 0$, it implies that $f'(x_T)<0$. Therefore, a necessary and sufficient condition for our bound in Theorem~\ref{theorem:FuCincreasing} to be tighter than $\frac{1}{p(x_\mathtt{PVT})}$ is that $x_Tf'(x_T)+\alpha f(x_T) < 0$ at $x_T = p^{-1}(\frac{\alpha}{\alpha+\gamma})$. Substituting $p(x_T) = \frac{\alpha}{\alpha+\gamma}$ into equation \eqref{eq:privatefoc11}, we obtain 
\begin{align*}
x_Tf'(x_T)+\alpha f(x_T) & = x_T\bar{r}'(x_T) \left[1-\frac{\alpha}{\gamma+\alpha}\right] + \alpha \bar{r}(x_T) - (\bar{r}(x_T)+k) (\gamma+\alpha) \frac{\alpha}{\gamma+\alpha}
\\ & = x_T\bar{r}'(x_T) \left[\frac{\gamma}{\gamma+\alpha}\right] + \alpha \bar{r}(x_T) - \alpha (\bar{r}(x_T)+k)
\\ & =  x_T\bar{r}'(x_T) \left[\frac{\gamma}{\gamma+\alpha}\right]  - k\alpha. 
\end{align*} 
The bound $(1+\frac{\gamma}{\alpha})$ is stronger if and only if 
\begin{align*}
& \bar{r}'\left(\left(\frac{\alpha}{\alpha+\gamma}\right)^{\frac{1}{\gamma}}\right) \leq  k\alpha \left(1+\frac{\alpha}{\gamma}\right) \left(1+\frac{\gamma}{\alpha}\right)^{\frac{1}{\gamma}}.
\end{align*}
When $\bar{r}(x_T)=ax_T+b$ and $\gamma=1$, we obtain $a \leq k(1+\alpha)^2$ as the necessary and sufficient condition for $(1+\frac{\gamma}{\alpha})$ to be the stronger bound. Note further that $\left(1+\frac{\gamma}{\alpha}\right)^{\frac{1}{\gamma}} \geq 1$ for every $\gamma > 1$, which gives us the second result.
\qed

\vspace{4mm}

\noindent \textbf{Proof of Proposition~\ref{proposition:socialwelfarealpha}.}
The effective rate of return function $f(x_T)$ is identical across all players under the conditions given in the proposition. Therefore, for any given $\alpha \in (0,1]$ and $x_T \in [0,1]$, $\Psi = \sum_{i=1}^{n} x_i^\alpha f(x_T)$ is maximized when $x_i = \frac{x_T}{n}$. When $x_T$ is nonzero, $\Psi$ must satisfy the first order condition of optimality in every player's investment $x_i$. Therefore, for every player $i$, we have 
\begin{align}
0 & = \frac{\partial \Psi}{\partial x_i} = \alpha x_i^{\alpha-1} f(x_T) + f'(x_T) \sum_{j=1}^{n} x_j^\alpha = \alpha \left(\frac{x_T}{n}\right)^{\alpha-1} f(x_T) + f'(x_T) n \left(\frac{x_T}{n}\right)^\alpha \nonumber
\\ & = \left(\frac{x_T}{n}\right)^{\alpha-1} [\alpha f(x_T) + x_T f'(x_T)]. \nonumber
\end{align}  
Thus the total investment at the social optimum, $x^\mathtt{OPT}_T$, must be a solution to $\alpha f(x_T) + x_T f'(x_T) = 0$. According to~\eqref{eq:pnexi}, this is the best response of a single player, $x_\mathtt{PVT}$, when the total investment by others is zero. As a result, $x^\mathtt{OPT}_T$ is unique from Lemma~\ref{lemma:CDuniquebr}. When the number of players is greater than 1, the total investment at the social optimum remains unchanged and each player invests an equal fraction of the total investment.
\qed

\section{Price of Anarchy}
\label{appendix:poa}

The strategies of the players at the PNE typically results in loss of welfare with respect to the social optimum. The term \textit{Price of Anarchy (PoA)}, first introduced by~\cite{koutsoupias1999worst}, is a widely used metric to quantify this loss of efficiency due to strategic behavior. We reproduce the definition here. 

\begin{definition}[~\cite{nisan2007algorithmic}]
Let $\mathbf{x}^\mathtt{OPT}$ be a strategy profile with optimal social welfare, and $\mathtt{NE}$ be the set of strategy profiles of the PNE in a strategic game. Then the Price of Anarchy (PoA) of the game is defined as
\begin{equation*}
\eta = \frac{\Psi(\mathbf{x}^\mathtt{OPT}), }{\min_{\mathbf{x}^* \in \mathtt{NE}} \Psi(\mathbf{x}^*)},
\label{eq:PoA}
\end{equation*}
where $\Psi(x_1,x_2,\ldots,x_n) = \sum_{i \in \mathcal{N}} \mathbb{E}(u_{i})$. It follows that $\eta \geq 1$.  
\end{definition} 

In the next result, we show that in Fragile CPR games with failure, the PoA is unbounded in the number of players when the players have homogeneous risk preferences. Thus greater competition leads to complete exhaustion of the utility obtained from the resource.

\begin{theorem} \label{theorem:PoAunbounded} Let $\{\Gamma_n\}_{n\geq2}$ be a sequence of Fragile CPR games with homogeneous players, and let $\eta_n$ be the price of anarchy of $\Gamma_n$. Then $\eta_n \to \infty$ as $n\to\infty$. \end{theorem}

\begin{proof}
Let the total investment at the PNE of $\Gamma_n$ be $x^*_{Tn}$. Recall from Proposition~\ref{proposition:socialwelfarealpha} that when players have homogeneous risk preferences, the total investment at the social optimum is independent of the number of players. We denote this socially optimal investment as $x^\mathtt{OPT}_T$, and we have that $x^\mathtt{OPT}_T = x_\mathtt{PVT}$, i.e., the optimal investment by a single player in the resource. Therefore, the optimal social welfare of $\Gamma_n$ is given by:
\begin{align*}
\Psi(\mathbf{x}^\mathtt{OPT}_n) &= f(x_\mathtt{PVT}) \sum_{i=1}^n \left(\frac{x_\mathtt{PVT}}{n}\right)^\alpha = f(x_\mathtt{PVT}) (x_\mathtt{PVT})^\alpha n^{1-\alpha}. 
\end{align*}  
The social welfare at the PNE with $n$ homogeneous players is 
\begin{align*}
\Psi(\mathbf{x}^*_n) &= \sum_{i=1}^{n} \left(\frac{x^*_{Tn}}{n}\right)^\alpha f(x^*_{Tn}) = n^{1-\alpha}(x^*_{Tn})^{\alpha}f(x^*_{Tn}) < n^{1-\alpha}\bar{y}^\alpha f(x^*_{Tn}).
\end{align*}
The inequality follows from Proposition~\ref{proposition:boundedpneXT}, which states that $\{x^*_{Tn}\}$ is a monotonically increasing sequence converging to $\bar{y}$ (defined in Lemma~\ref{lemma:interval}). Therefore, the ratio of $\Psi(\mathbf{x}^\mathtt{OPT}_n)$ and $\Psi(\mathbf{x}^*_n)$ is independent of the number of players $n$, and the denominator approaches $0$ as $n\to\infty$ (as $x^*_{Tn} \to \bar{y}$ and $f(\bar{y}) = 0$). 
\end{proof}

The theorem holds irrespective of the risk preferences of the players. The price of anarchy in games with heterogeneous players would depend on the respective risk attitudes, but as shown in the above theorem, it is not bounded in general.

\section{Proofs pertaining to fragility under heterogeneity}
\label{appendix:fuhproofs}

\noindent \textbf{Proof of Proposition~\ref{proposition:lossaversexi}.}
Let $x^*_T$ be the total investment of all the players in the PNE. Consider two players $i$ and $j$ such that $k_j \geq k_i$. Let their investments be $x^*_i$ and $x^*_j$, respectively. We have the following two exhaustive cases. 

\textbf{Case 1: $x^*_i=0.$}

From the arguments in Lemma~\ref{lemma:interval}, we know that $x^*_i=0$ if and only if $f_i(x^*_T) \leq 0$. Recall from~\eqref{eq:expectedutility} that the effective rate of return $f_i$ is strictly decreasing in $k_i$ for a fixed total investment in the CPR. As a result, we have $f_j(x^*_T) \leq 0$, and consequently $x^*_j=0$.

\textbf{Case 2: $x^*_i>0.$}

In this case, $x^*_i$ satisfies the necessary condition of optimality given in~\eqref{eq:pnexi}, i.e., 
\[ x^*_if'_i(x^*_T)+\alpha f_i(x^*_T)=0. \]
Thus, we have $f_i'(x^*_T)<0$. It follows from~\eqref{eq:effectiverormonotonicity} that 
\begin{align*}
f_j'(x^*_T) &= \bar{r}'(x^*_T)(1-p(x^*_T))-\bar{r}(x^*_T)p'(x^*_T)-k_jp'(x^*_T) \leq f'_i(x^*_T) < 0,
\end{align*}
since $k_j \geq k_i$ and $p'(x^*_T) > 0$. Assume on the contrary that $x^*_j>x^*_i$. We show that this results in a contradiction to~\eqref{eq:pnexi}. Formally, 
\begin{align*} 
x^*_jf'_j(x^*_T)+\alpha f_j(x^*_T) &\leq x^*_jf'_j(x^*_T)+\alpha f_i(x^*_T) < x^*_if'_i(x^*_T)+\alpha f_i(x^*_T) = 0.
\end{align*}
Therefore, in both cases we have $x^*_j \leq x^*_i$. 
\qed

\vspace{4mm}

\noindent \textbf{Proof of Theorem~\ref{theorem:fuh}.} Consider a Fragile CPR game $\Gamma_H \in \boldsymbol{\Gamma}$, $\Gamma_H \neq \Gamma_M$, i.e., the players in $\Gamma_H$ have heterogeneous loss aversion indices $\{k_1,k_2,\ldots,k_n\} \in \mathcal{K}$. Let the total investment at the respective PNEs of $\Gamma_H$ and $\Gamma_M$ be $x_{T,H}$ and $x_{T,M}$. Since the effective rate of return is positive for some $x_T \in [0,1]$ for a player with loss aversion index $k_M$, both $x_{T,H}$ and $x_{T,M}$ are nonzero. 

Denote the support of $\Gamma_H$ as $\mathcal{S}_H$, which contains the set of players with nonzero investments at the PNE of $\Gamma_H$. From Proposition~\ref{proposition:lossaversexi}, we know that players in the support have smaller loss aversion indices than those outside the support. Let $d \geq 1$ be most loss averse player in the support of $\Gamma_H$, i.e., $\mathcal{S}_H = \{1,2,\ldots,d\}$. In particular, player $1$ has a nonzero investment at the PNE of $\Gamma_H$. As a result, $x_{T,H} \in \mathcal{I}_1$, where $\mathcal{I}_1$ is the interval defined in Lemma~\ref{lemma:interval} for player $1$. In other words, $f_1(x_{T,H}) > 0$ and $f'_1(x_{T,H}) < 0$.  

Recall from~\eqref{eq:expectedutility} that $f_i(x_T)$ and $f'_i(x_T)$ can be written as,
\begin{align*}
f_i(x_T) &:= u(x_T) -k_i p(x_T), \text{   and   }
\\ -f'_i(x_T) &:= -u'(x_T) + k_ip'(x_T),
\end{align*}
where $u(x_T)=\bar{r}(x_T)(1-p(x_T))$. 

Since $p(x_T)$ is strictly increasing, $f'_i(x_{T,H}) < 0$ for any player $i$ with $k_i \geq k_1$. Furthermore, let $\bar{k} > k_1$ be the smallest loss aversion index such that $u(x_{T,H}) -\bar{k} p(x_{T,H}) \leq 0$. Accordingly, we have $k_d < \bar{k}$ and $k_{d+1} \geq \bar{k}$. 

Let the investment by player $i$ at the PNE of $\Gamma_H$ be denoted as $x_{i,H}$. From \eqref{eq:focbestresponse}, we know that for a player $i \in \mathcal{S}_H$, $x_{i,H} = g_i(x_{T,H}) = \frac{\alpha f_i(x_{T,H})}{-f'_i(x_{T,H})} > 0$. For a player $j \notin \mathcal{S}_H$, $k_j \geq \bar{k}$ and therefore, $f_j(x_{T,H}) \leq 0$. Accordingly, we have the following characterization of $x_{T,H}$:
\begin{equation}\label{eq:fuh_defineh}
x_{T,H} = \sum^n_{i=1} \max(g_i(x_{T,H}),0) = \sum^n_{i=1} \max\left(\alpha \frac{u(x_{T,H}) -k_i p(x_{T,H})}{-u'(x_{T,H}) + k_ip'(x_{T,H})},0\right) \triangleq \sum^n_{i=1} \max(h_{x_{T,H}}(k_i),0).
\end{equation}
In other words, $h_{x_{T,H}}(k)$ is a function of $k$ with parameter $x_{T,H}$. It is continuous in $k \in [k_1,\infty)$ from the above discussion. We now show that $h_{x_{T,H}}(k)$ is convex in $k$ for $k \in [k_1,\bar{k})$, which is sufficient for the convexity of $\max(h_{x_{T,H}}(k_i),0)$ for $k \in [k_1,nk_M]$. 

First note that for $k\in[k_1,\bar{k})$, we have $u(x_{T,H}) - k p(x_{T,H}) > 0$ and $-u'(x_{T,H}) + kp'(x_{T,H}) > 0$. Thus, we obtain
\begin{equation}\label{eq:convexh1}
u(x_{T,H})p'(x_{T,H}) > k p(x_{T,H})p'(x_{T,H}) > u'(x_{T,H})p(x_{T,H}),
\end{equation}
for $k\in[k_1,\bar{k})$. We now compute the first and second derivatives of $h_{x_{T,H}}(k)$. For notational convenience, we denote $u(x_{T,H}),u'(x_{T,H}),p(x_{T,H})$ and $p'(x_{T,H})$ as $u,u',p$ and $p'$ respectively. For $k\in[k_1,\bar{k})$, we obtain
\begin{align*}
\frac{d h_{x_{T,H}}(k)}{d k} & = \alpha \frac{(-u'+kp')(-p)-(u-kp)p'}{(-u'+kp')^2} = \alpha \frac{u'p-up'}{(-u'+kp')^2} < 0, \text{  and   }
\\ \frac{d^2 h_{x_{T,H}}(k)}{d k^2} & = \alpha \frac{-2(u'p-up')p'}{(-u'+kp')^3} > 0;
\end{align*}
both inequalities follow from \eqref{eq:convexh1}. From the convexity of $\max(h_{x_{T,H}}(k_i),0)$ for $k \in [k_1,nk_M]$, we obtain
\begin{equation}\label{eq:fuh_contradiction}
\frac{x_{T,H}}{n} = \frac{1}{n} \sum^n_{i=1} \max(h_{x_{T,H}}(k_i),0) \geq \max(h_{x_{T,H}}(k_M),0) = \max(g_{M}(x_{T,H}),0),
\end{equation}
following Jensen's inequality. Here $g_{M}(x_T)$ is as defined in \eqref{eq:focbestresponse} for a player with $k=k_M$. Note that since $k_M > k_1$, we have $f'_M(x_{T,H}) < 0$. We have the following two cases depending on the value of $g_{M}(x_{T,H})$. 

Suppose $g_{M}(x_{T,H}) \leq 0$. Since $f'_M(x_{T,H}) < 0$, we must have $f_M(x_{T,H}) \leq 0$. As a result, for any $x_T \in (x_{T,H},1]$, we have $f_M(x_{T}) < 0$. Since $x_{T,M} > 0$ at the PNE of $\Gamma_M$ from our assumptions, we must have $x_{T,M} \leq x_{T,H}$.

Now suppose $g_{M}(x_{T,H}) > 0$. Assume on the contrary that $x_{T,H} < x_{T,M}$. Since $f'_M(x_{T,H}) < 0$, we must have $f_M(x_{T,H}) > 0$. As a result, $[x_{T,H},x_{T,M}] \subseteq \mathcal{I}_M$, where $\mathcal{I}_M$ is the interval defined in Lemma~\ref{lemma:interval} for a player with $k=k_M$. Then from Lemma~\ref{lemma:garg}, we have $g_{M}(x_{T,H}) > g_{M}(x_{T,M})$. However, from \eqref{eq:fuh_contradiction}, we obtain,
$$ \frac{x_{T,H}}{n} \geq g_{M}(x_{T,H}) > g_{M}(x_{T,M}) = \frac{x_{T,M}}{n},$$
which is a contradiction. This concludes the proof. 

\qed

\vspace{4mm}

\noindent \textbf{Proof of Proposition~\ref{proposition:alphauniformfragility}}
Let $k_i$ be the loss aversion index of the players in $\Gamma_i$, $i\in\{1,2\}$, and let $f_i$ denote the effective rate of return of the players in the respective games. Let $k_1 < k_2$ as required by the proposition. Then from the definition of the effective rate of return function, we know that $f_2(x_T) < f_1(x_T)$ and $f'_2(x_T) < f'_1(x_T)$ for $x_T \in [0,1]$. Let $x^*_{Ti} \in \mathcal{I}_i$ be the total investment at the PNE of $\Gamma_i$, $i\in\{1,2\}$, where $\mathcal{I}_i$ is the interval defined in Lemma~\ref{lemma:interval} for a player in $\Gamma_i$. Assume on the contrary that $x^*_{T1} \leq x^*_{T2}$. 

Note that since $x^*_{T1} \in \mathcal{I}_1$, we have $f'_1(x^*_{T1}) < 0$. From the discussion above, it follows that $f'_2(x^*_{T1}) < 0$. Since $x^*_{T1} \leq x^*_{T2}$, we therefore have $f'_2(x_{T}) < 0$ for $x_T \in [x^*_{T1}, x^*_{T2}]$. On the other hand, $f_2(x^*_{T2}) > 0$ since $x^*_{T2} \in \mathcal{I}_2$. Therefore, we must have $f_2(x^*_{T1}) \geq f_2(x^*_{T2}) > 0$, and as a result $x^*_{T1} \in \mathcal{I}_2$. Thus, we have
\begin{align*}
\frac{x^*_{T2}}{n} = \frac{\alpha f_2(x^*_{T2})}{(-f'_2(x^*_{T2}))} \leq & \frac{\alpha f_2(x^*_{T1})}{(-f_2'(x^*_{T1}))} < \frac{\alpha f_1(x^*_{T1})}{(-f_1'(x^*_{T1}))} = \frac{x^*_{T1}}{n}. 
\end{align*}
The equalities follow from the first order condition of optimality (equation~\eqref{eq:homogeneous_foc_PNE}). The first inequality is a consequence of Lemma~\ref{lemma:garg} as $x^*_{Ti} \in \mathcal{I}_2, i \in \{1,2\}$, while the second inequality follows from the monotonicity of $f$ and $f'$ in the loss aversion index as mentioned above. Thus we have the desired contradiction, and as a result, we must have $x^*_{T1} > x^*_{T2}$.
\qed

\end{document}